\documentclass[journal,onecolumn,12pt]{IEEEtran}
\usepackage{amsmath,amssymb,amsfonts,mathrsfs}
\usepackage{graphicx}
\usepackage{xcolor}
\usepackage{amsthm}
\usepackage{amsmath}
\usepackage{bbm}
\usepackage{stix}
\usepackage{algorithm}
\usepackage{algpseudocode}
\usepackage{hyperref}
\usepackage{verbatim} 
\usepackage{float}
\newtheorem{theorem}{Theorem}
\newtheorem{lemma}[theorem]{Lemma}
\newtheorem{proposition}[theorem]{Proposition}
\newtheorem{corollary}[theorem]{Corollary}

\usepackage{amsmath,blkarray,booktabs,bigstrut}
\usepackage{tikz}
\usepackage{color}
\usepackage{color, colortbl}
\definecolor{Gray}{gray}{0.9}
\usepackage{caption}
\theoremstyle{definition}
\newtheorem{definition}[theorem]{Definition}
\newtheorem{remark}[theorem]{Remark}
\newtheorem{example}[theorem]{Example}

\usepackage{blkarray}
\usepackage{nomencl}
\makenomenclature
\renewcommand{\nomname}

\ifCLASSINFOpdf

\else
 
\fi

\begin{document}

\title{The Schur product of evaluation codes and its application to CSS-T quantum codes and private information retrieval}

\author{
    \c{S}eyma Bodur\textsuperscript{1}, 
    Fernando  Hernando\textsuperscript{2}, 
    Edgar Mart\'inez-Moro\textsuperscript{1}, 
    Diego Ruano\textsuperscript{1} \\
    \textsuperscript{1}IMUVA-Mathematics Research Institute, Universidad de Valladolid, Spain
    \textsuperscript{2}Departamento de Matem{\'a}ticas \& Institut Universitari de Matem{\`a}tiques i Aplicacions de Castell{\'o}, Universitat Jaume I, Spain \\
    \thanks{Emails: \{seyma.bodur, edgar.martinez, diego.ruano\}@uva.es, carrillf@uji.es. }

\thanks{This work was supported in part by Grants PID2022-138906NB-C21 and PID2022-138906NB-C22 funded by MICIU/AEI/10.13039/501100011033 and by ERDF/EU, and by Grant CONTPR-2019-385 funded by Universidad de Valladolid and Banco Santander.}
}
\markboth{}{}

\maketitle

\begin{abstract}
In this work, we study the Schur (componentwise) product of monomial–Cartesian codes by exploiting its correspondence with the Minkowski sum of their defining exponent sets.  We show that $ J$-affine variety codes are well suited for such products, generalizing earlier results for cyclic, Reed–Muller, hyperbolic, and toric codes.  Using this correspondence, we construct CSS–T quantum codes from weighted Reed–Muller codes and from binary subfield-subcodes of $J$-affine variety codes, leading to codes with better parameters than previously known.  Finally, we present Private Information Retrieval (PIR) constructions for multiple colluding servers based on hyperbolic codes and subfield-subcodes of $ J$-affine variety codes, and show that they outperform existing PIR schemes.
\end{abstract}


\IEEEpeerreviewmaketitle

\section{Introduction}

The Schur (componentwise) product of linear codes over a finite field has emerged as a fundamental operation in both classical and quantum coding theory \cite{hugues}.  Cyclic, Reed–Muller, hyperbolic, and toric codes have all been used to compute Schur products, leading to applications in coding theory and cryptography.  These families belong to the class of evaluation codes, obtained by evaluating univariate or multivariate polynomials on finite point sets.  In this paper, we focus on two principal applications: CSS–T quantum codes and private information retrieval (PIR).

In Section \ref{se:dos}, we extend the computations of \cite{Cascudo,prod,prodtoric} to monomial–Cartesian codes, defined by evaluating monomials on affine varieties.  We prove that $J$-affine variety codes (see \cite{Jaffine15,LCD,Jaffine17}) are well suited for componentwise multiplication and give an explicit Minkowski-sum description of their Schur product, thus generalizing earlier results for cyclic, Reed–Muller, hyperbolic, and toric codes.  We further show that taking subfield-subcodes commutes with these products, enabling the construction of binary evaluation codes with strong multiplication properties.
    
Section \ref{se:tres} addresses CSS–T codes \cite{csst1,csst2}, i.e.\ Calderbank–Shor–Steane constructions that admit a transversal T gate.  These codes extend classical CSS codes-which correct both bit-flip and phase-flip errors-by enabling fault-tolerant implementation of the non-Clifford T gate, a key ingredient for universal quantum computation.  Algebraically, a CSS–T code is defined by a pair of binary linear codes $(C_1,C_2)$ satisfying $C_2  \subseteq C_1 \cap  (C_1^{\star2})^\perp$  \cite{csst-us}. We show that binary weighted Reed–Muller codes and the binary subfield-subcodes of $J$–affine variety codes also yield CSS–T codes, and for the same length and minimum distance, achieve strictly larger dimension than the existing constructions.

In Section \ref{sec:PIR}, we turn to PIR schemes with multiple servers. A PIR protocol enables a user to retrieve a record from a coded distributed database without revealing its index to any of the $t$ servers, even if they collude \cite{PIR-RM}. In this setting, the PIR rate is defined as the ratio between the size of the requested file and the total downloaded information, and the capacity of a PIR scheme is the supremum of all achievable PIR rates \cite{capacit}. In \cite{CapacityCodedPIR}, the case of non-colluding servers and MDS codes is considered, and their capacity is studied. In \cite{PIRCodedColluding}, a PIR scheme with colluding servers using Generalized Reed-Solomon codes is presented; its rate is determined by the minimum distance of the Schur product of the storage and retrieval codes, and it is shown to be asymptotically capacity-achieving. Reference \cite{TowardCapacity} considers the non-asymptotic capacity case and establishes that PIR schemes from MDS codes are capacity-achieving. However, when constructing PIR schemes from MDS codes, one must work over a large base field. For practical purposes, binary or small fields may be preferable in some scenarios, and in \cite{PIR-RM}, a scheme based on binary Reed–Muller codes is proposed. Using the same approach, \cite{PIRcyclic} presents a PIR scheme based on cyclic codes. In this section, we consider the framework of \cite{PIR-RM} and \cite{PIRcyclic}, that is, PIR schemes over a binary field or over a small field compared to the code length. Therefore, we will compare our constructions only with PIR schemes in this framework. Thus, following the transitive-code approach of \cite{PIR-RM}, we prove that decreasing monomial–Cartesian codes (Definition~\ref{de:decreasing}) and $J$-affine variety codes defined via complete consecutive cyclotomic sets (Lemma~\ref{lem:jafin-trans}) admit a transitive automorphism group, and thus support PIR schemes with adjustable privacy levels.  First, we show that hyperbolic codes outperform classical Reed–Muller–based PIR in the rate–privacy trade-off.  Then, by employing subfield-subcodes of $J$-affine codes, we obtain PIR schemes that achieve improved PIR rate at fixed privacy, outperforming known protocols based on Reed–Muller, cyclic, and Berman codes \cite{PIR-RM,PIRcyclic,BermanCodes}. We consider both single-variable and multivariable settings and provide explicit parameter comparisons. We note that our constructions also yield binary PIR schemes.
 
We also comment on potential extensions to secure multi-party computation (MPC), which enables multiple parties to jointly compute a function while preserving both input privacy and output correctness.  Multiplicative secret-sharing schemes based on linear codes require a large code dimension, a large minimum distance of the dual code, and a high minimum distance of the code's componentwise square \cite{SMSSS}.  As noted in Example~\ref{le:SSDual}, the dual of the subfield-subcode of the componentwise product of two $J$–affine variety codes does not, in general, coincide with the componentwise product of their dual subfield-subcodes. We leave the precise conditions under which this equality holds to future work.

\section{(Affine variety) Monomial Cartesian codes and $J$-affine variety codes}\label{se:dos}

Let $\mathbb{F}_q$ be the finite field with $q$ elements, a power of the prime $p$. For $j=1,\ldots, m$, let $Z_j \subseteq \mathbb{F}_q$, with $z_j = \# Z_j \ge 2$. We consider the set $Z =Z_1  \times \cdots \times Z_m \subseteq \mathbb{F}_q^m$. Consider now $E_j = \{0,1, \ldots , z_j -1 \}$, for $j=1,\ldots, m$, and  $E = E_1 \times \cdots \times E_m$. Note that $n=\#Z=\#E = \prod_{j=1}^m z_j$. For any $\boldsymbol{e}=(e_1,\ldots, e_m) \in E$, set $X^{\boldsymbol{e}} = X_1^{e_1} \cdots X_{m}^{e_m}$.

Let $I$ be the ideal of $\mathbb{F}_{q}[X_1,\dots,X_m]$ generated by the polynomials 
$Q_j(X_j)= \prod_{\beta \in Z_j}(X_j - \beta)$,  for $j=1, \ldots , m$. Note that the field equations $X_j^q -X_j \in I$, for all $j=1, \ldots, m$. We consider the quotient ring $R:=\mathbb{F}_{q}[X_1,\dots,X_m] /I$, then any class of polynomials has a unique representative of the form  $$f(X_1,\dots,X_m)=\sum_{\boldsymbol{e}\in E} f_{\boldsymbol{e}}X^{\boldsymbol{e}}=\sum_{(e_1,\dots,e_m)\in E} f_{e_1,\dots,e_m}X_1^{e_1}\cdots X_m^{e_m},$$with $f_{e_1,\dots,e_m}\in\mathbb{F}_{q}$. Abusing notation, we will refer to $f$ as a polynomial in $R$. Let the zero locus of the ideal $I$ be equal to $Z=\{\boldsymbol{P}_1, \ldots, \boldsymbol{P}_n \}$, with the points of $Z$ ordered in a specific way. Consider the linear evaluation map given by
$$
\operatorname{ev}_{Z}\colon R \longrightarrow \mathbb{F}_{q}^{n} \textrm{, } \quad 
\operatorname{ev}_{Z}(f)=\left(f(\boldsymbol{P}_i)\right)_{i=1,\ldots,n}.
$$ It is well-defined and bijective. For $\Delta \subseteq E$, the \textit{(Affine variety) Monomial Cartesian code} (\textit{MCC}) $C_{\Delta,Z}$ is the image of $\{f \in R \mid \operatorname{supp}(f)\subseteq \Delta\}$ under the evaluation map $\operatorname{ev}_{Z}$, that is, $$C_{\Delta,Z}:=\operatorname{Span}\{\operatorname{ev}_{Z}(X^{\boldsymbol{e}}) \mid \boldsymbol{e}\in \Delta\} \subseteq \mathbb{F}_{q}^{n}.$$Since the evaluation map is injective, the dimension of $C_{\Delta,Z}$ is equal to the cardinality of $\Delta$.

The minimum distance of $C_{\Delta,Z}$, $d(C_{\Delta,Z})$, can be estimated mainly in two different ways. The first way is by the foot-print bound \cite{footprint}, which follows from considering a Gr{\"o}bner basis of the ideal $I$ \cite{CLO}:\begin{equation}
   d(C_{\Delta,Z}) \ge \delta_{\operatorname{FB}}(C_{\Delta,Z}) = \min \left\{ \prod_{i=1}^m (z_j-e_j) : \boldsymbol{e} \in \Delta  \right\}. 
\end{equation}
The second one is considering the multiplicative nature of the minimum distance of monomial Cartesian codes \cite{soprunov,dmult}. Let $\Delta \subseteq \Delta_1 \times \cdots \times \Delta_m$, then \begin{equation}\label{eq:distanceProduct}d(C_{\Delta,Z}) \ge \prod_{i=1}^m d(C_i), \end{equation}
where $C_i = C_{\Delta_i,Z_i}$ is a uni-variate evaluation code. 
Hyperbolic codes \cite{hyperbolic}, and in general monomial-decreasing Cartesian codes (see definition \ref{de:decreasing}),  are optimal for the foot-print bound \cite{prod}. For the multiplicative bound,  there are optimal codes coming from $\Delta$ sets that are a Cartesian product ($\Delta = \Delta_1 \times \cdots \times \Delta_m$) \cite{cartesian}. We introduce now these families of codes.

Consider $s \ge 0$ and let $\Delta = \{ \boldsymbol{e} \in \{0,\ldots, q-1 \}^m : e_1 + \cdots + e_m \le s    \}$ and $Z = \mathbb{F}_q^m$, then the code $C_{\Delta,Z}$ is a  \emph{Reed-Muller code} and it is denoted by $\mathrm{RM}_q(s,m)$. The minimum distance of a Reed-Muller code is equal to $(q-b)q^{m-1-a}$, where $s=a(q-1)+b$, with $0\le b \le q-1$ \cite{dRM}. Let $s, s_1, \ldots, s_m > 0$, without loss of generality, assume that $s_1  \le \ldots \le s_m $. Set $\Delta = \left\{ \boldsymbol{e} \in \{0,\ldots, q-1 \}^m : s_1e_1 + \ldots + s_m e_m \leq s\right\}$. Then, the code $C_{\Delta,Z}$ is a \emph{weighted Reed-Muller}, with $\mathcal S = \left( s_1, \ldots, s_m \right)$, and we denote it by $\mathrm{WRM}_q(s,m, \mathcal S)$. If $s_1 = \ldots = s_m = 1$, then the $\mathrm{WRM}_q(s,m, \mathcal S)$ is a Reed-Muller code, $\mathrm{RM}_q(\lfloor s \rfloor,m)$.
Finally, let $s \ge 0$ and set $\Delta = \{ \boldsymbol{e} \in \{0,\ldots, q-1 \}^m : (q - e_1) \cdots  (q-e_m) \ge s    \}$ and $Z = \mathbb{F}_q^m$, then the code $C_{\Delta,Z}$ is known as a \emph{hyperbolic code} \cite{hyperbolic}.
The foot-print bound is sharp for these families of codes. Actually, one has that the footprint bound is sharp if and only if all the elements $(\beta_1,\ldots, \ldots, \beta_m)$ with $0\le \beta_i \le \alpha_i$ belong to $\Delta$, where $\delta_{\operatorname{FB}}(C_{\Delta,Z}) = \prod_{i=1}^m (q-\alpha_i)$ \cite{prod}, these codes are known as monomial-decreasing Cartesian codes.

This paper aims to investigate the componentwise product of monomial Cartesian codes and its applications. The \emph{componentwise product, or Schur, or star product}, of two vectors $\boldsymbol{\alpha},\boldsymbol{\beta} \in \mathbb{F}_q^n$ is equal to $\boldsymbol{\alpha} \star \boldsymbol{\beta} = (\alpha_1 \beta_1, \ldots , \alpha_n \beta_n)\in \mathbb{F}_q^n$. Given two codes $C_1, C_2 \subseteq \mathbb{F}_q^n$, their componentwise product, or Schur, or star product, is equal to $$C_1 \star C_2= \operatorname{Span}_{\mathbb{F}_q}\{\boldsymbol{c}_1 \star\boldsymbol{c}_2 \mid  \boldsymbol{c}_1 \in C_1, \boldsymbol{c}_2 \in C_2\}.$$ We denote $C^{\star 2}=C\star C.$ The \emph{Minkowski sum} of two sets $A,B \subset \mathbb{N}^m$ is equal to $$A + B = \{\boldsymbol{a}+\boldsymbol{b} = (a_1 + b_1, \ldots, a_m + b_m)  \mid \boldsymbol{a} \in A, \boldsymbol{b} \in B\}.$$There is a strong connection between the componentwise product of evaluation codes and the Minkowski sum. Indeed, in one direction one has that $\operatorname{ev}_{Z}(f_1) \star \operatorname{ev}_{Z}(f_2) = \operatorname{ev}_{Z} (f_1 f_2)$, if $f_1 f_2  \in E$. However, on the other side if $f_1 f_2 \notin E$, then we have to reduce this polynomial by the ideal $I$ and this reduction is not explicit, in general. Actually, for   $\mathbf e\notin E$, the evaluation $\operatorname{ev}_{Z}(X^\mathbf e)$ is equivalent to $\operatorname{ev}_{Z}(h)$ where  $X^\mathbf e \equiv h \mod  I$ and the $\operatorname{supp}(h)\subseteq E$. Nevertheless,  we do not know  the $\operatorname{supp}(h)$ for the reduced polynomial of $X^\mathbf e$ modulo $I$, in general.

However, the product of polynomials can sometimes be explicitly reduced by $I$, that is, we may reduce the  Minkowski sum of $\Delta_1 + \Delta_2$ in $E$, as can be found in the literature for the componentwise product of evaluation codes. For instance, this is the case when $Z=(\mathbb{F}^\ast_q)^m$, $E=\{0,\ldots,q-2\}^m$, and the codes are called  \emph{toric codes} \cite{ruanotoric2,prodtoric}. In this later case of codes, the ideal $I$ is generated by $X_j^{q-1} - 1$, for $j=1, \ldots, m$. Hence, for $\mathbf e \in \mathbb{N}^m$ we have that $\operatorname{ev}_{Z} (X^\mathbf e) = \operatorname{ev}_Z (X^{\overline{\mathbf e}})$, where  $\overline{\mathbf e} = (e_1 \mod q-1 , \ldots , e_m \mod q-1)$ and the reduction modulo $q-1$ is taken in $\{0,\ldots, q-2\}$. That is, ${\overline{\mathbf e}} \in  \mathbb{Z}_{q-1}^m = \{0,\ldots, q-2\}^m$. Therefore, if we define $\overline{A}=\{ \overline{\mathbf a} \mid \mathbf a \in A\} \subseteq E$, for $A \subset \mathbb{N}^m$, then $C_{\Delta_1,(\mathbb{F}^\ast_q)^m} \star C_{\Delta_2,(\mathbb{F}^\ast_q)^m}$ is equal to $C_{\overline{\Delta_1 + \Delta_2},(\mathbb{F}^\ast_q)^m}$, where $\overline{\Delta_1 + \Delta_2} \subseteq E$. 

Another known case is $Z=\mathbb{F}_q^m$ and $E=\{0,\ldots,q-1\}^m$. For instance, for  Reed-Muller and Hyperbolic codes \cite{prod}. Now the reduction is a bit more involved but still explicit. The ideal $I$ is generated by the field equations $X_j^{q} - X_j$, for $j=1, \ldots, m$. Hence, for $\mathbf e \in \mathbb{N}^m$ we have that $\operatorname{ev}_{Z} (X^\mathbf e) = \operatorname{ev}_Z (X^{\overline{\mathbf e}})$, where  $\overline{\mathbf e} = (e_1 \mod q-1 , \ldots , e_m \mod q-1)$ and the reduction modulo $q-1$ is taken in $\{1,\ldots, q-1\}$ if $e_i >0$ and it is not reduced otherwise. For instance, $0 \mod q-1 =0$, $q-1 \mod q-1 =q-1$, and $q \mod q-1 = 1$. Thus,  ${\overline{\mathbf e}} \in  \{\{0\} \cup \mathbb{Z}_{q-1}\}^m = \{\{0\} \cup \{1, \ldots , q-1\}\}^m$. If we define now $\overline{A}=\{ \overline{\mathbf a} \mid \mathbf a \in A\} \subseteq E$, for $A \subset \mathbb{N}$. And we have that $C_{\Delta_1,\mathbb{F}_q^m} \star C_{\Delta_2,\mathbb{F}_q^m}$ is equal to $C_{\overline{\Delta_1 + \Delta_2},\mathbb{F}_q^m}$, where $\overline{\Delta_1 + \Delta_2} \subseteq E$.

\subsection{$J$-Affine variety codes}

This paper aims to consider (Affine variety) Monomial Cartesian codes where one can find an explicit reduction of $\Delta_1 + \Delta_2$ in $E$. That is, we extend the previous two cases considered in the literature \cite{prodtoric,prod}. We propose to consider $J$-affine variety codes, and their subfield subcodes, that have been successfully used to obtain quantum codes \cite{Jaffine15,Jaffine17}, and LCD codes \cite{LCD}, with excellent parameters. This family of codes considers sets $Z$ with a cyclic structure that is well posed for considering subfield subcodes, and that, moreover, they can also be useful for computing their componentwise product. The results concerning the componentwise product in this section are new.

Let us introduce $J$-affine variety codes. Fix $m$ integers $N_j>1$ such that $N_j-1$ divides $q-1$ for $j=1, \ldots, m$. Consider a subset $J \subseteq \{1, \ldots, m\}$ and the ideal $I_J$ in $\mathcal{R}$ generated by the binomials $X_j^{N_j} - X_j$ when $j \not \in J$ and by $X_j^{N_j -1} - 1$ otherwise. We have that $Z_j$ is the zero locus of the corresponding binomial, and we have hence defined the set $Z$. Note that the $j$-th coordinate, for $j \in J$, of the points in $Z_J$ is different from zero, and the length is given by $n_J = \prod_{j \notin J} N_j \prod_{j \in J} (N_j -1)$. Moreover, denote $T_j = N_j -2$  when $j \in J$ and $T_j = N_j -1$ otherwise, then define  
$$
E_J = \{0,1,\ldots,T_1\}\times \cdots\times\{0,1,\ldots,T_m\},
$$which agrees with the definition of the set $E$ at the beginning of this section. Then, from this definition of $Z$ (and $E$) we have a particular class of (Affine variety) Monomial Cartesian codes known as \emph{$J$-affine variety codes}. In particular, for $N_j =q$, we recover the two previous situations for $J=\{1,\ldots , m \}$  (toric case), and $J=\emptyset$ ($Z=\mathbb{F}_q^m$ case). We will consider a simpler notation for $J$-affine codes that ease the reading and strengths the dependence on the set $J$: the quotient ring will be denoted by $R_J=R$, the evaluation map will be denoted by $\mathrm{ev}_J = \mathrm{ev}_Z$ and the {\it $J$-affine variety code given by $\Delta \subseteq E$ } is the $\mathbb{F}_q$-vector subspace $C^J_\Delta$ of $\mathbb{F}_q^{n_J}$ generated by $\mathrm{ev}_J (X^{\boldsymbol{a}})$, $\boldsymbol{a} \in \Delta$.    
The dual code can be computed using the following result \cite{Jaffine15}.

\begin{proposition}
\label{pr:dualJ1}
Let $J \subseteq \{ 1 , 2, \ldots , m\}$, consider $\boldsymbol{a}, \boldsymbol{b} \in \mathcal{H}_J$ and let $X^{\boldsymbol{a}}$ and $X^{\boldsymbol{b}}$ be two monomials representing elements in $\mathcal{R}_J$. Then, $\mathrm{ev}_J ( X^{\boldsymbol{a}}) \cdot \mathrm{ev}_J (X^{\boldsymbol{b}})$ is different from $0$ if,  and only if, the following two conditions are satisfied.
\begin{itemize}
\item For every $j \in J$, it holds that $a_j + b_j \equiv  0 \mod (N_j -1)$, (i.e.,  $a_j = N_j -1 - b_j$ when $a_j  + b_j > 0$ or $a_j=b_j=0$).
\item For every $j \notin J$, it holds that \begin{itemize}
\item either $a_j  + b_j > 0$ and $a_j + b_j \equiv 0 \mod (N_j -1)$,  (i.e.,  $a_j = N_j -1 - b_j$  if $0 < a_j, b_j < N_j -1$ or $(a_j,b_j) \in \left\{(0,N_j -1), (N_j -1,0), (N_j -1,N_j -1)  \right\}$ otherwise),
\item or $a_j = b_j = 0$ and $p \not | ~ N_j$.
\end{itemize}
\end{itemize}
\end{proposition}

If we set $ E' := E_{\{1,2, \ldots, m\}}$ and pick $\Delta \subseteq E_\emptyset$, define  $\Delta^\perp$ as
$$ E_J \setminus \{ (N_1 -1 -a_1, N_2 -1 - a_2, \ldots, N_m- 1 -a_m) \; | \; \boldsymbol{a} \in \Delta\},$$
if $\Delta \subseteq E'$. When $\Delta \not\subseteq E'$ define $\Delta^\perp$ as
$$
E_J\setminus \left( \{ (N_1 -1 -a_1, N_2 -1 -a_2, \ldots , N_m -1 -a_m ) | \boldsymbol{a} \in \Delta \cap E' \} \cup \{ \boldsymbol{a}' | \boldsymbol{a} \in \Delta , \boldsymbol{a} \notin E' \} \right),$$
where we set $a'_j = N_j -1 -a_j$ if $a_j \neq N_j -1$ and $a'_j$ equals either $N_j -1$ or $0$ otherwise.

Next, we state the result about the dual code and self-orthogonality of a $J$-affine code that follows from the previous result. 

\begin{proposition}
\label{pr:dualJ2}
With notations as above, let $\Delta$ be a subset of $E_J$. Then $\left(C_\Delta^J\right)^\perp = C^J_{\Delta^\perp}$ whenever $\Delta \subseteq E'$. Otherwise, it holds that $\left(C_\Delta^J\right)^\perp \subseteq C^J_{\Delta^\perp}$.
\end{proposition}

 For $\Delta_1, \Delta_2 \subseteq E_J$, we will consider their Minkowski sum $\Delta_1 + \Delta_2 \subset \mathbb{N}^m$ and then reduce it in $E_J$ in the following way. For $\mathbf a \in \mathbb{N}^m$  we define $\overline{\mathbf a} = (\overline{a}_1 , \ldots , \overline{a}_m)\in E_J$, where, if $j\in J$, $\overline{a}_j = a_j \mod N_j -1$, that is  $\overline{a}_j \in \{0, \ldots, N_j -2\} = \{0,1, \ldots, T_j\}$ . Otherwise, if $j\notin J$, then $\overline{a}_j$ is equal to $0$ if $a_j = 0$, and it is equal to $a_j \mod N_J -1$, where the reduction modulo $N_j-1$ is taken in $\{1,\ldots N_J -1\}$, i.e. $\overline{a}_j \in \{ 0 ,1, \ldots, N_j -1\} = \{0,1, \ldots , T_j\}$. Furthermore,  $\operatorname{ev}_{Z} (X^\mathbf a) = \operatorname{ev}_Z (X^{\overline{\mathbf a}})$, since we evaluate classes of polynomials modulo the ideal $I_J$ in $\mathcal{R}$ generated by the binomials $X_j^{N_j} - X_j$ when $j \not \in J$ and by $X_j^{N_j -1} - 1$ otherwise. Thus, for $A \subset \mathbb{N}^m$, we define $\overline{A} =\{ \overline{\mathbf a} \mid \mathbf a\in A\} \subseteq E_J$.  Thus, the componentwise product of $J$-affine variety codes is given by the following result.

\begin{theorem}\label{th:prodJafines}
Let $N_j>1$ such that $N_j-1$ divides $q-1$ for $j=1, \ldots, m$ and $J \subseteq \{1, \ldots, m\}$. Let $\Delta_1, \Delta_2 \subseteq E_J$. Then, the componentwise product of $C^J_{\Delta_1}$ and $C^J_{\Delta_2}$ is given by $$C^J_{\Delta_1} \star C^J_{\Delta_2} = C^J_{\overline{\Delta_1+\Delta_2}}.$$  
\end{theorem}

\begin{proof}
The result follows from the previous discussion and the fact that  $\operatorname{ev}_{Z}(f_1) \star \operatorname{ev}_{Z}(f_2) = \operatorname{ev}_{Z} (f_1 f_2)$ and that $\operatorname{ev}_{Z} (f) = \operatorname{ev}_{Z} (f \mod I)$. \end{proof}
Notice how this extends the computations in \cite{prodtoric,prod} (for $N_j=q$ for all $j$).

\subsection{Subfield subcodes of $J$-Affine codes}\label{se:subf}

Given a linear code $C$ of lenght $n$ over $\mathbb{F}_ {q}$ and $\mathbb{F}_{q'} \subseteq \mathbb{F}_{q}$, the subfield-subcode over $\mathbb{F}_{q'}$ is $S(C)=C\cap \mathbb{F}_{q'}^n$. i.e., the set of codewords in $C$ with all the coordinates over the subfield $\mathbb{F}_ {q'}$.  

We are going to consider subfield subcodes of $J$-Affine codes. A subset ${I}$ of the Cartesian product $E_J$ is called a cyclotomic set with respect to $p$ if $p \cdot \boldsymbol{x}\in{I}$ for every element $\boldsymbol{x}=(x_1,\ldots,x_m)\in{I}$, where we define $p \cdot \boldsymbol{x}=(p x_1,\ldots,p x_m)$. A cyclotomic set ${I}$ is said to be minimal (with respect to $p$) if it consists exactly of all elements expressible as $p^i \cdot \boldsymbol{x}$ for some fixed element $\boldsymbol{x}\in{I}$ and some nonnegative integer $i$. In the case of one variable, they are usually called cyclotomic cosets. 

Within each minimal cyclotomic set ${I}$, we select a representative element $\boldsymbol{a}=(a_1,\ldots,a_m)$ consisting of nonnegative integers as follows: first, $a_1$ is chosen as the minimum first coordinate among all nonnegative representatives of elements in ${I}$; next, $a_2$ is the minimum second coordinate among those elements having first coordinate equal to $a_1$; and similarly, we define coordinates $a_3,\ldots,a_m$. We denote by ${I}_{\boldsymbol{a}}$ the cyclotomic set with representative $\boldsymbol{a}$, and by $\mathcal{A}$ the set of representatives of all minimal cyclotomic sets. Thus, the set of minimal cyclotomic sets is given by  $\{ {I}_{\boldsymbol{a}}\}_{\boldsymbol{a} \in \mathcal{A}}$. In addition, we denote its cardinality by $i_{\boldsymbol{a}} = \mathrm{card}({I}_{\boldsymbol{a}})$.

The subfield-subcode associated to a given $J$-affine variety code $C_\Delta^J$ over the finite field $\mathbb {F}_{q'}=\mathbb {F}_{p^r}$ is defined as: $$C_\Delta^{J,\sigma} = S(C_\Delta^J) = C_\Delta^J \cap \mathbb{F}_{p^r}^{n_J}.$$

Consider now the following maps:  $\mathrm{tr}: \mathbb{F}_q \rightarrow \mathbb{F}_p$, $\mathrm{tr}(x) = x + x^p + \cdots + x^{p^{r-1}}$;  $\mathbf{tr}: \mathbb{F}_q^{n_J} \rightarrow \mathbb{F}_p^{n_J}$ given componentwise by $\mathrm{tr}(x)$, and $\mathcal{T}: \mathcal{R}_J \rightarrow \mathcal{R}_J$ defined by $\mathcal{T} (f) = f + f^p + \cdots + f^{p^{r-1}}$.  

The dimension of $C_\Delta^{J, \sigma}$ is given in \cite[Theorem 11]{LCD}. Note that, when computing the dimension, only those cyclotomic sets that are complete will contribute, that is, when a cyclotomic set $I_{\boldsymbol{a}} \subseteq \Delta$. The minimum distance of $C_\Delta^{J,\sigma}$ is lower bounded by the minimum distance of $C_\Delta^J$.

\begin{theorem}
\label{ddimension}
Let $\Delta$ be a subset of $\mathcal{H}_J$ and set $\xi_{\boldsymbol{a}}$ a primitive element of the field $\mathbb{F}_{p^{i_{\boldsymbol{a}}}}$. Then, a basis of the vector space $C_\Delta^{J,\sigma}$ is given by the images under the map $\mathrm{ev}_J$ of the set of classes in $R_J$
$$\bigcup_{ \boldsymbol{a} \in \mathcal{A}| {I}_{\boldsymbol{a}} \subseteq \Delta
} \left\{ \mathcal{T}_{\boldsymbol{a}} (\xi_{\boldsymbol{a}}^{s} X^{\boldsymbol{a}}) | 0 \leq s \leq i_{\boldsymbol{a}} -1 \right\},$$and the dimension of $C_\Delta^{J,\sigma}$ is given by the cardinality of $\Delta^\sigma := \cup_{\boldsymbol{a} \in \mathcal{A}| \mathfrak{I}_{\boldsymbol{a}} \subseteq \Delta}  {I}_{\boldsymbol{a}}$
\end{theorem}

Computing the componentwise product of subfield subcodes can be tricky, as the next example shows.

\begin{example}
Let $q=4$, $\alpha$ a primitive element ($\alpha^3=1$), and  subfield-subcodes over $\mathbb{F}_2$. Let $C_1$ and $C_2$ the linear codes generated by $(1,\alpha,0)$, and $(0,\alpha^2,1)$, respectively. One can easily check that $S(C_1)=S(C_2)=\{(0,0,0)\}$, but $S(C_1 \star C_2)$ is generated by $(0,1,0)$.
\end{example}

However, the componentwise product of the subfield subcodes of $J$-affine variety codes can be explicitely computed if we consider their defining set to be a union of complete cyclotomic cosets.  

\begin{lemma}
Let $N_j>1$ such that $N_j-1$ divides $q-1$, for $j=1, \ldots, m$, and $J \subseteq \{1, \ldots, m\}$. Let $\Delta_1, \Delta_2 \subseteq E_J$ be a union of complete cyclotomic cosets. Then, the componentwise product of $C^{J,\sigma}_{\Delta_1}$ and $C^{J,\sigma}_{\Delta_2}$ is given by $$S(C_{\Delta_1}^J \star C^J_{\Delta_2}) = C^{J,\sigma}_{\Delta_1} \star C^{J,\sigma}_{\Delta_2} = C^{J,\sigma}_{\overline{\Delta_1+\Delta_2}}.$$ 
\end{lemma}

\begin{proof}
By Theorem \ref{th:prodJafines}, it follows that $S(C_{\Delta_1}^J \star C^J_{\Delta_2}) = S (C^{J}_{\overline{\Delta_1+\Delta_2}} )  = C^{J,\sigma}_{\overline{\Delta_1+\Delta_2}}$.

On the other hand, under the asumption that $\Delta_1$ and $\Delta_2$ are complete cyclotomic cosets, reasoning as before Theorem \ref{th:prodJafines}, one has that $S(C^J_{\Delta_1}) \star S(C^J_{\Delta_2}) = C^{J,\sigma}_{\Delta_1}  \star  C^{J,\sigma}_{\Delta_2}$ is equal to $C^{J,\sigma}_{\overline{\Delta_1+\Delta_2}}$, and the result holds.
\end{proof}

As we mentioned in the introduction, for having applications to multi-party computation one would have to deal with both the subfield subcode of both the componentwise square of an evaluation code and its dual. The next example shows that it is not possible, even though one works with a union of complete cyclotomic cosets, since the equality $S(C_1 \star C_2)^\bot = S(C_1)^\bot \star S(C_2)^\bot$ does not hold, in general.

\begin{example}\label{le:SSDual}
 Let $N_1=16$ and $J=\{1\}$. Consider $\Delta_1=\{1,2,4,8\}$ and $\Delta_2=\{0\}$, both a union of complete cyclotomic cosets. One has that $$\Delta_1^{\perp}=\{0,1,2,3,4,5,6,8,9,10,12\}, \mbox{~and~} \Delta_2^{\perp}=\{1,2,3,4,5,6,7,8,9,10,11,12,13,14\}.$$Then the codes $C^J_{\Delta_1}$ and $C^J_{\Delta_2}$ have length 15 and are defined over $\mathbb{F}_{16}$. We consider subfield subcodes over $\mathbb{F}_{2}$. Moreover, notice that $\Delta_1+\Delta_2=\Delta_1$.

 Then, $S(C^J_1\star C^J_2)^{\perp}=S(C^J_1)^{\perp}$, but $S(C^J_{\Delta_1})^{\perp} \star S(C^J_{\Delta_2})^{\perp}$ is equal to $S(C^J_\Delta)$, with $\Delta= \Delta_1^{\perp}+\Delta_2^{\perp}=\{0,1,\ldots,14\}$. Thus we have that 
$$S(C^J_{\Delta_1}\star C^J_{\Delta_2})^{\perp}\ne S(C^J_{\Delta_1})^{\perp} \star S(C^J_{\Delta_2})^{\perp}.$$

\end{example}

\section{CSS-T codes}\label{se:tres}

We follow the convention of using $[[n,k,d]]$ to denote a quantum code encoding $k$ qubits (known as logical qubtis) into $n$ physical qubits and that can correct less than $d$ erasures. We consider the CSS construction \cite{CSS1,CSS2}.

\begin{theorem}\label{css}
Let $C_1, C_2 \subseteq \mathbb{F}_{2}^n$ be linear codes with dimension $k_1, k_2$, respectively, and such that $C_2\subseteq C_1$. Then, there is an $[[n,k_1-k_2,d]]$ quantum code with 
$$
d=\min \left\{\mathrm{wt} \left(C_1\setminus C_2 \right), \mathrm{wt}\left(C_2^\perp\setminus C_1^\perp\right) \right\},
$$where $\mathrm{wt}$ denotes the minimum Hamming weight.
\end{theorem}

CSS-T codes are a class of CSS codes that may implement the $T$ gate transversally, which is a crucial step for achieving fault-tolerant quantum computation. They may reduce the overhead associated with magic state distillation, a common technique used to implement non-Clifford gates in fault-tolerant quantum circuits. CSS-T codes were introduced in \cite{csst1,csst2} and they were algebraically characterized in terms of the Schur product of the pair of binary classical linear codes that define them in \cite{csst-us}, note that this definition specifically requires binary codes. More concretely, they are defined from a pair of binary linear codes $(C_1,C_2)$, called a CSS-T pair, such that $$C_2 \subseteq C_1 \cap (C_1^{\star 2})^\bot.$$ Moreover, we have that for a  CSS-T pair $(C_1, C_2)$ then $\min \{\mathrm{wt}(C_1),\mathrm{wt}(C_2^\perp)\}=\mathrm{wt}(C_2^\perp)$, and the parameters of the corresponding CSS-T code are (\cite[Corollary 2.5]{csst-us}) $$[[n,k_1-k_2,\ge \mathrm{wt}(C_2^\perp)]].$$

\subsection{CSS-T codes from Weighted Reed-Muller codes}

CSS-T codes arising from Reed-Muller codes were considered in \cite{csst-RM} and from cyclic codes in \cite{csst-us}. It was shown that the parameters of CSS-T codes coming from cyclic codes may outperform the ones coming from Reed-Muller codes. In this work, we show that we can define CSS-T codes from weighted Reed-Muller codes and that their parameters can improve the parameters in \cite{csst-RM,csst-us}. Specifically, we will consider that $C_1$ is a weighted Reed-Muller code and $C_2$ is a Reed-Muller code (a Reed-Muller code being a weighted Reed-Muller code with trivial weights).

Initially, we require a lemma concerning the inclusion or nesting of Weighted Reed-Muller codes within Reed-Muller codes from \cite{WRM}. Note that it is required in \cite{WRM} that the weights $\mathcal S = \left( s_1, \ldots, s_m \right)$ for defining a weighted Reed-Muller code verify $s_1 \le \cdots \le s_m$. Without loss of generality, we assume that the weights are ordered in this work as well.

\begin{lemma}\label{rem:weightedreedmullercode}
   One has that
    \[
\mathrm{RM}(v_{\min}(s), m) \subseteq \mathrm{WRM}(s, m,\mathcal S) \subseteq \mathrm{RM}(v_{\max}(s), m),
\]
where $v_{\min}(s)= \max  \{ v \mid s \geq \sum \limits_{i=m-v+1}^{m}s_i \}$, $v_{\max}(s)= \max \{ v \mid s \geq \sum \limits_{i=1}^{v}s_i  \}$.
\end{lemma}

We can now state the main result of this section.
  
\begin{theorem}
For $m \ge 2$, let $C_1$ be binary weighted Reed-Muller $C_1=\mathrm{WRM}(s,m,\mathcal S)$ and $C_2$ be the binary Reed-Muller code $C_2=\mathrm{RM}(r,m)$, with $r\le\max \{ v \mid s \geq \sum \limits_{i=m-v+1}^{m}s_i\}$. Then $(C_1,C_2)$ is a CSS-T pair if $a +r < m$, where $a = max \{ j \mid 2s \geq \displaystyle \sum_{i=1}^{ j}s_i \}$.

     The parameters of the associated CSS-T quantum code are $[[2^m, k_1-k_2,  2^{r+1} ]]$, where $k_1= \dim (C_1)$, and $k_2= \dim (C_2) = \sum_{i=0}^{s} \binom{m}{i}$, 
    \end{theorem}
    \begin{proof}

%\dim (C_1) = |\{ (i_1,i_2,\cdots,i_m) : s_1i_1+\cdots+s_mi_m \le s  \}|$
 
From Lemma \ref{rem:weightedreedmullercode},  it follows that $C_2 \subseteq C_1$ because $r\le\max \{ v \mid s \geq \sum \limits_{i=m-v+1}^{m}s_i\}$.

Note that in general, the Schur square of a weighted Reed-Muller code is not a weighted Reed-Muller code. However, we have the inclusion
   \[  \mathrm{WRM}(s,m,\mathcal S)^{\star 2} \subseteq \mathrm{WRM}(2s,m,\mathcal S). \] 
   %\subset RM(a,m),$$where $a = max \{ j | 2s \geq \sum\limits_{i=1}^{j}s_i \}$.

Combining this fact with Lemma~\ref{rem:weightedreedmullercode}, we can ensure that \begin{equation}\label{eq:proWRM}
   \mathrm{WRM}(s,m,\mathcal S)^{\star 2}  \subseteq \mathrm{RM}(a,m).
  \end{equation} 

The condition $C_2 \subseteq (C_1^{\star 2})^\bot$ translates to the inclusion 

\[   \mathrm{RM}(r,m) \subseteq  (\mathrm{WRM}(s,m,\mathcal S)^{\star 2} )^\bot ,  \]

which in turn is equivalent to 

\[  \mathrm{WRM}(s,m,\mathcal S)^{\star 2}   \subseteq   \mathrm{RM}(r,m)^\bot = \mathrm{RM} (m-r-1,m). \]

Furthermore, by combining the previous equation with Equation (\ref {eq:proWRM}), we have that $C_2 \subseteq (C_1^{\star 2})^\bot$ if 

\[ \mathrm{RM}(a,m) \subseteq \mathrm{RM}(m-r-1,m),\]

which holds if $a \le m-r-1$. The parameters follow from \cite[Corollary 2.5]{csst-us}. This completes the proof.
    \end{proof}

\begin{example}
    Consider the binary weighted Reed-Muller code $C_1=\mathrm{WRM}(5,7,(1,2,2,2,2,2,2))$  with the set $\Delta=\{(i_1,i_2,i_3,i_4,i_5,i_6,i_7) : i_1+2i_2+2i_3+2i_4+2i_5+2i_6+2i_7 \leq 5 \}$, which has parameters $[128,44,16]$, and the binary Reed-Muller code $C_2= \mathrm{RM}(1,7)$ with parameters $[128,8,64]$. We have that $k((C_1^{\star 2})^{\bot})=11$ and  $C_2 \subseteq C_1 \cap (C_1^{\star 2})^{\bot}$, as established by the previous result. Thus, $(C_1 , C_2)$ is a CSS-T pair whose associated CSS-T code has parameters $[[128, 36,4]]$. This construction allows us to generate further examples, listed in Table \ref{table:VII}. These examples outperform the CSS-T codes presented in \cite{csst-RM,csst-us}, as shown in table \ref{table:jcss-t}.

\begin{table}[h!]
\begin{center}
\begin{tabular}{|c|c|c|c|c|c|} 
\hline
$\mathbf{C_2}$ & $\mathbf{C_1}$ & $\mathbf{C_1^{\star 2}}$ & $\mathbf{(C_1^{\star 2})^{\bot}}$ & $\mathbf{C_2^{\bot}}$ & CSS-T\\ \hline  
$ \mathbf{[128,8,64]}$ & $ \mathbf{[128,44,16]}$ & $ \mathbf{[128,117,4]}$ &  $ \mathbf{[128,11,32]}$ & $ \mathbf{[128,120,4]}$ & $\mathbf{[[128,36,4]]}$\\  
$ \mathrm{RM}(1,7)$ & $ \mathrm{WRM}(5,7, (1,2,2,2,2,2,2))$ & & & &\\ \hline 
$ \mathbf{[256,37,64]}$ & $ \mathbf{[256,58,32]}$ & $ \mathbf{[256,198,8]}$ &  $ \mathbf{[256,58]}$ & $ \mathbf{[256,219,8]}$ & $\mathbf{[[256,21,8]]}$\\  
$ \mathrm{RM}(2,8)$ & $ \mathrm{WRM}(5,8, (1,2,2,2,2,2,2,2))$ & & & &\\ \hline 
$ \mathbf{[512,10,128]}$ & $ \mathbf{[512,186]}$ & $ \mathbf{[512,494]}$ &  $ \mathbf{[512,18]}$ & $ \mathbf{[512,502,4]}$ & $\mathbf{[[512,176,4]]}$ \\  
$ \mathrm{RM}(1,9)$ & $ \mathrm{WRM}(7,9, (1,2,2,2,2,2,2,2,2))$ & & & &\\ \hline 
$ \mathbf{[1024,56,128]}$ & $ \mathbf{[1024,260]}$ & $ \mathbf{[1024,932]}$ &  $ \mathbf{[1024,92]}$ & $ \mathbf{[1024,968,8]}$ & $\mathbf{[[1024,204,8]]}$ \\
$ \mathrm{RM}(2,10)$ & $ \mathrm{WRM}(7,10, (1,2,2,2,2,2,2,2,2,2))$  & & &&\\ \hline
\end{tabular}
\end{center}
\caption{Parameters of CSS-T codes from Weighted Reed-Muller Codes $C_1$  and Reed-Muller Codes $C_2$}\label{table:VII}
\end{table}

\end{example}

\subsection{CSS-T codes from subfield subcodes of $J$-affine codes}

 Moreover, we can extend the previous result to $J$-affine codes, thereby increasing the range of possible parameters, particularly the length. This extension allows us to explore a broader constellation of codes, offering wider flexibility in code design.

\begin{theorem}\label{theo:css'}
Let $q$ be a power of 2 and the sets $\Delta_1, \Delta_2 \subseteq E' \subset  E_J$ be a union of complete cyclotomic cosets. The pair of binary codes $(C_{\Delta_1}^{J,\sigma}, C_{\Delta_2}^{J,\sigma}) $ is a CSS-T pair if and only if
    \begin{enumerate}
        \item $\Delta_2 \subseteq \Delta_1$, and
        \item $\Delta_1 + \Delta_1 \subseteq \Delta_2^\bot $
    \end{enumerate}
The parameters of the associated CSS-T quantum code are $$\left[\left[ \prod\limits_{j\notin J} N_j \prod\limits_{j\in J} (N_j-1), \# \Delta_1 - \# \Delta_2, \geq \mathrm{wt}((C_{\Delta_2}^{J,\sigma})^\bot)=\mathrm{wt}(C_{\Delta_2^\bot}^{J,\sigma})\right]\right].$$
\end{theorem}

\begin{proof}
By Proposition \ref{pr:dualJ2}, $\left(C_{\Delta_i}^{J, \sigma}\right)^\perp = C^{J, \sigma}_{\Delta_i^\perp}$, for $i=1,2$, because $\Delta_1, \Delta_2 \subseteq E' \subset E_J$ and they are a union of complete cyclotomic cosets. Hence, $\Delta_2 \subseteq \Delta_1$ if and only if $C_{\Delta_2}^{J,\sigma} \subseteq C_{\Delta_1}^{J,\sigma}$. Moreover, since $\Delta_1 + \Delta_1 \subseteq \Delta_2^\bot $, then  $(C_{\Delta_1}^{J,\sigma})^{\star 2} \subseteq (C_{\Delta_2}^{J,\sigma})^\bot$ which is equivalent to $C_{\Delta_2}^{J,\sigma} \subseteq ((C_{\Delta_1}^{J,\sigma})^{\star 2})^\bot$. The parameters follow from \cite[Corollary 2.5]{csst-us} and the result holds.
\end{proof}

The previous result is neat; however, the hypothesis $\Delta_1, \Delta_2 \subseteq E'$ is impractical. The following one allows us to consider $\Delta_1, \Delta_2 \subseteq E_J$, that is, without the restriction $\Delta_1, \Delta_2 \subseteq E'$. Thus it is more flexible. However, in this case, one must be careful when computing the dual codes, as equality no longer holds by Proposition \ref{pr:dualJ2}, but rather only a containment.

\begin{theorem} \label{theo:css}
Let $q$ be a power of 2 and the sets $\Delta_1, \Delta_2  \subseteq E_J$ be a union of complete cyclotomic cosets. The pair of binary codes $(C_{\Delta_1}^{J,\sigma}, C_{\Delta_2}^{J,\sigma}) $ is a CSS-T pair if 
    \begin{enumerate}
        \item $\Delta_2 \subseteq \Delta_1$, and
        \item For all $\mathbf{a} \in \overline{\Delta_1 + \Delta_1 + \Delta_2}$ there exists $j\in \{1,\ldots,m\}$ such that, 
        \begin{itemize}
            \item $a_j \neq 0$, if $j \in J$,
            \item $a_j \neq N_j -1$, if $j \notin J$.
        \end{itemize}
    \end{enumerate}
The parameters of the associated CSS-T quantum code are $$\left[\left[ \prod\limits_{j\notin J} N_j \prod\limits_{j\in J} (N_j-1), \# \Delta_1 - \# \Delta_2, \geq \mathrm{wt}((C_{\Delta_2}^{J,\sigma})^\bot)\right]\right].$$
\end{theorem}
\begin{proof} 

Since $\Delta_2 \subseteq \Delta_1$, it follows that $C_{\Delta_2}^{J,\sigma} \subseteq C_{\Delta_1}^{J,\sigma}$.

By Proposition \ref{pr:dualJ2}, $\left(C_{\Delta_i}^{J,\sigma}\right)^\perp \subseteq  C^{J,\sigma}_{\Delta_i^\perp}$, for $i=1,2$, because $\Delta_1, \Delta_2 \subseteq E' \subset E_J$ and they are unions of complete cyclotomic cosets. 

Observe that  $\mathbb{1}=(1\ldots,1)=\operatorname{ev}_{Z}(x_1^0\cdots x_m^0)$, so $\mathbf{b}=\mathbf{0}$  in Proposition \ref{pr:dualJ1}, we may conclude  applying it that $\mathbb{1}=(1\ldots,1) \in \left(C_{\Delta}^J\right)^\bot$ if and only if for all $\mathbf{a} \in \Delta$ there exists $j\in \{1,\ldots,m\}$ such that , 
        \begin{itemize}
           \item $a_j \neq 0$, if $j \in J$,
            \item $a_j \neq N_j -1$, if $j \notin J$. Notice that case  $a_j=0$ is excluded because  $p=2 \mid N_J$.
        \end{itemize}

Furthermore, observe that  $C^{J,\sigma}_{\Delta_2} \subseteq ((C^{J,\sigma}_{\Delta_1})^{\star 2})^\bot$ if and only if $\mathbb{1} \in (  C^{J,\sigma}_{\Delta_1} \star C^{J,\sigma}_{\Delta_1}  \star  C^{J,\sigma}_{\Delta_2} )^\bot =  (C^{J,\sigma}_{\overline{\Delta_1 + \Delta_1 + \Delta_2}} )^\bot$. This is supported by the previous claim and the hypothesis of the result. The parameters follow from \cite[Corollary 2.5]{csst-us}, and the result holds.
   \end{proof}

Based on the previous results, we propose now a concrete construction of subfield subcodes of $J$-Affine variety codes that yield excellent families of CSS-T quantum codes. First, we will consider a one-variable subfield subcode of a $J$-affine variety code, then we will extend it to an $m$-variable $J$-affine variety code

Let $C^1 = C^{J,\sigma}_{\Delta^1}$ and $C^2 = C^{J,\sigma}_{\Delta^2}$ a pair of one-variable subfield subcodes with  $N_1-1\mid 2^r-1$ and $J=\emptyset$ (we evaluate at zero). Assume that $(C^1, C^2)$ is a CSS-T pair whose associated CSS-T code has parameters $[[N_1,\#\Delta^1-\#\Delta^2, d]]$.

We consider now the extension to  $m$ variables. Let $(N_i-1) \mid 2^r - 1$, for $i=2,\ldots m$, and $1\le m_1\le m$. Let $J = \{m_1+1,\ldots,m\}$, that is, we evaluate at the zeros of  
$\prod_{j=1}^{m_1} (x_j^{N_j} -x_j ) \prod_{j=m_1+1}^{m} (x_j^{N_j-1} - 1)$. Set $\Delta_1$ to be 
$$\Delta_1=\Delta^1 \times \{0,1,\ldots,N_2-1\}\times \cdots \times \{0,1,\ldots,N_{m_1}-1\}\times \{0,1,\ldots,N_{m_1+1}-2\}\times \cdots \times\{0,1,\ldots,N_m-2\}.$$ 

Let $C \subseteq \mathbb{F}_{q^r}^n$, with $n=\prod\limits_{j\notin J} N_j \prod\limits_{j\in J} (N_j-1)$, be a hyperbolic code with designed minimum distance $d$ and let ${\Delta_H}$ such that $C^J_{{\Delta_H}} = C^\bot$. Now define $\Delta_2$ to be the $\cup_{\boldsymbol{a}\in{\Delta_H}} I_{\mathbf{a}}$, i.e., for each element in ${\Delta_H}$ we append to $\Delta_2$ the whole cyclotomic coset $I_{\mathbf{a}}$. Thus, $\Delta_2$ is a union of complete cyclotomic cosets.

Let $C_i = C^{J,\sigma}_{\Delta_i}$, for $i=1,2$. Since $\Delta^2\subset\Delta^1$ then $\Delta_2\subset\Delta_1$ and therefore $C_2 \subseteq C_1$. Moreover, since by construction $N_1-1\notin \overline{\Delta^1 + \Delta^1 + \Delta^2}$, it follows that 
there is no $\mathbf{a} \in \overline{\Delta_1 + \Delta_1 + \Delta_2}$ whose first coordinate, $a_1$, is equal to $N_1-1$. Thus, from Theorem \ref{theo:css} we have that $(C_1,C_2)$ is a CSS-T pair whose associated quantum CSS-T codes has parameters given by the next result.

\begin{corollary}\label{th:CSSTjafines3}
Let  $q=2^r$ and $(N_i-1) \mid 2^r - 1$, for $i=2,\ldots m$. Let $J = \{m_1+1,\ldots,m\}$ where $1\le m_1\le m$.  Consider the construction of  $\Delta_1$ and $\Delta_2$ designed before, then  there exist a CSS-T codes with parameters$$\left[\left[\prod_{j=1}^{m_1} N_j\prod_{j=m_1+1}^m (N_j-1),\#\Delta_1-\#\Delta_2,d\right]\right].$$
\end{corollary}

We consider now several examples of the construction given in Corollary \ref{th:CSSTjafines3}. The notation is as before. All the cyclotomic sets used to define the codes in Examples \ref{ex:jcsst128} and \ref{ex:jcsst192}, and Table \ref{table:jcss-t} are given explicitly in Table \ref{table:cyclotomic-cosets-jcss-t}.

\begin{example} \label{ex:jcsst128}
We consider Corollary~\ref{th:CSSTjafines3} in the case of three variables with $J = \emptyset$, i.e., $m = m_1 = 3$, and $r=4$. Let $N_1 = 16$, $N_2 = 4$, and $N_3 = 2$. Then, the length of the CSS-T code is $n = N_1 \cdot N_2  \cdot N_3 = 128$.

Let $\Delta^1 = \Delta^2 = I_0 \cup I_1 = \{0, 1, 2, 4, 8\}$. Since $\Delta^2$ contains three consecutive integers, the BCH bound implies that
\[
d\left((C^{J,\sigma}_{\Delta^2})^\perp\right) \geq 4.
\]

As described above, we define
\[
\Delta_1 = \Delta^1 \times \{0,1,2,3\} \times \{0,1\}, 
\]
and
\[
\Delta_2 = I_{(0,0,0)} \cup I_{(1,0,0)} \cup I_{(0,1,0)} \cup I_{(0,0,1)}.
\]
Note that $\Delta_2 \subseteq \Delta_1$, $\#\Delta_1 = 5 \cdot 4 \cdot 2 = 40,$ and $\#\Delta_2 = 1 + 4 + 2 + 1 = 8$.

To estimate the minimum distance of $(C^{J,\sigma}_{\Delta_2})^\perp$, we observe that it is a subcode of a hyperbolic code with minimum distance 4, and hence
\[
d\left((C^{J,\sigma}_{\Delta_2})^\perp\right) \geq 4.
\]

Therefore,  there exists a CSS-T  code with parameters
\[
[[ 128,\ 40 - 8 = 32,\ d \geq 4 ]].
\]
\end{example}

\begin{example} \label{ex:jcsst192}
We consider now Corollary~\ref{th:CSSTjafines3} in the case of two variables, with $J = \{2\}$, and $r=6$. Hence, $m = 2$ and $m_1 = 1$. Let $N_1 = 64$ and $N_2 = 4$. Then, the length of the CSS-T code is $n = N_1(N_2 - 1)  = 192$.

Let $\Delta^1 = I_0 \cup I_1 \cup I_3 \cup I_5  \cup I_9$, and $\Delta^2 = I_0 \cup I_1 = \{0,1,2,4,8,16,32\}$. Again, by the BCH bound
\[
d\left((C^{J,\sigma}_{\Delta^2})^\perp\right) \geq 4.
\]

Now, let
\[
\Delta_2 = I_{(0,0)} \cup I_{(1,0)} \cup I_{(0,1)},
\]
which has cardinality $\#\Delta_2 = 9$. Since the dual code $(C^{J,\sigma}_{\Delta_2})^{\perp}$ is a subcode of the corresponding hyperbolic code of distance $4$, it follows that
\[
d((C^{J,\sigma}_{\Delta_2})^{\perp}) = 4.
\]

Finally, let
\[
\Delta_1 = \Delta^1 \times \{0,1,2\},
\]
with cardinality $\#\Delta_1 = 66$.

Therefore, there exists a CSS-T code with parameters
\[
[[192,\ 66 - 9 = 57,\ d \geq 4 ]].
\]

In Table~\ref{table:cyclotomic-cosets-jcss-t}, we list the selections of $\Delta_1$ and $\Delta_2$ for other values of $N_2$. For instance, replacing $N_2-1=3$ with $N_2-1=7$ yields a quantum code with parameters $[[448,141,4]]$, and replacing it with $N_2-1=9$ yields one with parameters $[[576,183,4]]$.

\end{example}

\begin{table}[h]
\begin{center}
\resizebox{\textwidth}{!}{%
\begin{tabular}{|c|c|c|c|c|c|c|}
\hline
$ \Delta_2 $  & $\Delta_1$ & CSS-T   \\\hline
$I_{(0,0,0)}\cup I_{(1,0,0)}\cup I_{(0,1,0)}\cup I_{(0,0,1)}$  & $\{0,1,2,4,8\} \times  \{0,1,2,3\} \times \{0,1\} $ &  $[[128,32,4]]$  \\ \hline

$I_{(0,0)}\cup I_{(1,0)}\cup I_{(0,1)}$  & $\{0,1,2,4,8,16,32,3,6,12,24,48,33,5,10,20,40,17,34,9,18,36\} \times  \{0,1,2\}  $ &  $[[192,57,4]]$  \\ \hline
$ I_{( 0, 0)} \cup I_{( 0,  1 )}\cup I_{( 0, 1 )} \cup I_{( 1, 1 )}\cup I_{( 3, 0 )}\cup I_{( 5, 0 )}$ & $\{0,1,2,4,8,16,32,64,3,6,12,24,48,96,65,5,10,20,40,80,33,66,9,18,36,72,17,34,68\} \times \{0,1\} $ &  $[[256,28,8]]$  \\ \hline
$I_{(0,0)}\cup I_{(1,0)}\cup I_{(0,1)}$  & $\{0,1,2,4,8,16,32,3,6,12,24,48,33,5,10,20,40,17,34,9,18,36\} \times  \{0,1,2,3,4,5,6\}  $ &  $[[448,141,4]]$  \\ \hline
$ I_{( 0, 0, 0,0 )} \cup I_{( 0, 0, 0,1 )}\cup I_{( 0, 0, 1,0 )}\cup I_{( 0, 1, 0,0 )}  \cup I_{( 1, 0, 0,0 )}$ & $  \{0,1,2,4,8,16,32,3,6,12,24,48,33,5,10,20,40,17,34,9,18,36\} \times \{0,1\}\times \{0,1\}\times \{0,1\} $ &  $[[512,166,4]]$  \\ \hline
$I_{(0,0)}\cup I_{(1,0)}\cup I_{(0,1)}$  & $\{0,1,2,4,8,16,32,3,6,12,24,48,33,5,10,20,40,17,34,9,18,36\} \times  \{0,1,2,3,4,5,6,7,8\}  $ &  $[[576,183,4]]$  \\ \hline
$ I_{( 0, 0 )} \cup I_{( 0, 1 )}  \cup I_{( 1, 0 )} \cup I_{( 1,1 )}\cup I_{( 3,0 )}$& $   \{0, 1, 2, 4, 8, 16, 32, 64, 128, 256, 3, 6, 12, 24, 48, 96, 192, 384, 257, 5, 10, 20, 40, 80, 160, 320, 129, 258, 7, 14, 28, 56, 112, 224, 448, 385, 259, 9, 18, 36, 72,$ &\\ & $144, 288, 65, 130, 260, 11, 22, 44, 88, 176, 352, 193, 386, 261, 13, 26, 52, 104, 208, 416, 321, 131, 262, 17, 34, 68, 136, 272, 33, 66, 132, 264, 19, 38, 76, 152, 304,$&\\ & $ 97, 194, 388, 265, 21, 42, 84, 168, 336, 161, 322, 133, 266, 25, 50, 100, 200, 400, 289, 67, 134, 268, 35, 70, 140, 280, 49, 98, 196, 392, 273, 37, 74, 148, 296, 81, 162, 324, 137,$& $[[1024,231,6]]$\\ & $ 274, 41, 82, 164, 328, 145, 290, 69, 138, 276, 73, 146, 292  \} \times \{0,1\} $&     \\ \hline
$ I_{( 0, 0 )} \cup I_{( 0, 1 )}  \cup I_{( 1, 0 )} \cup I_{( 1,1 )}\cup I_{( 3,0 )}\cup I_{( 5,0 )}$& $   \{0, 1, 2, 4, 8, 16, 32, 64, 128, 256, 3, 6, 12, 24, 48, 96, 192, 384, 257, 5, 10, 20, 40, 80, 160, 320, 129, 258, 7, 14, 28, 56, 112, 224, 448, 385, 259, 9, 18, 36, 72,$ &\\ & $144, 288, 65, 130, 260, 11, 22, 44, 88, 176, 352, 193, 386, 261, 13, 26, 52, 104, 208, 416, 321, 131, 262, 17, 34, 68, 136, 272, 33, 66, 132, 264, 19, 38, 76, 152, 304,$&\\ & $ 97, 194, 388, 265, 21, 42, 84, 168, 336, 161, 322, 133, 266, 25, 50, 100, 200, 400, 289, 67, 134, 268, 35, 70, 140, 280, 49, 98, 196, 392, 273, 37, 74, 148, 296, 81, 162, 324, 137$,& $[[1024,222,8]]$\\ & $ 274, 41, 82, 164, 328, 145, 290, 69, 138, 276, 73, 146, 292  \} \times \{0,1\} $&     \\ \hline
\end{tabular}
}
\end{center}
\caption{Cyclotomic cosets used in Examples \ref{ex:jcsst128} and \ref{ex:jcsst192}, and Table \ref{table:jcss-t}.} \label{table:cyclotomic-cosets-jcss-t}
\end{table}

To conclude this section, Table~\ref{table:jcss-t} compares our codes with those in \cite{csst-RM,csst-us}. The length and minimum distance of the codes in each row coincide, while our codes have larger dimension. Note also that the CSS-T codes from the $J$-affine variety construction outperform those from the WRM construction; however, for length 128 the WRM  CSS-T code surpasses the $J$-affine one. A heuristic procedure to increase the dimension of a CSS-T code without reducing its minimum distance was proposed in \cite[Corollary~3.9]{csst-us}. The resulting codes are labeled ``Improved Reed–Muller'' and ``Improved Extended Cyclic'' in Table~\ref{table:jcss-t}. We have not applied this heuristic, so there remains potential to enhance our parameters using \cite[Corollary~3.9]{csst-us}.

\begin{table}[H]
\begin{center}
\begin{tabular}{|c|c|c|c|c|c|c|}
\hline
$n$ & Reed-Muller & Improved Reed-Muller & Extended Cyclic & Improved Extended Cyclic & WRM & $J$-Affine \\ \hline  
$128$ & $[[128,21,4 ]]$ &$[[128,26,4]] $ & $[[128,28,4]]$ &  & $\mathbf{[[128,36,4]]}$ & $\mathbf{[[128,32,4]]}$ \\ \hline 
$256$ &  & & $[[256,20,8]]$ & $[[256,22,8]]$ &$\mathbf{[[256,21,8]]} $ & $\mathbf{[[256,28,8]]}$ \\ \hline 
$512$ & $[[512,120,4]]$ &$[[512,133,4]]$ & $[[512,147,4]]$  & $[[512,148,4]]$  &$\mathbf{[[512,176,4]]}$ & $\mathbf{[[512,166,4]]}$ \\ \hline 
$1024$ & & & $[[1024,210,6]]$  & $[[1024,217,6]]$  & & $\mathbf{[[1024,231,6]]}$  \\ \hline 
$1024$ &$[[1024,120,8]]$ & $[[1024,125,8]]$& $[[1024,190,8]]$ & $[[1024,192,8]]$  &$\mathbf{[[1024,204,8]]}$ & $\mathbf{[[1024,222,8]]}$  \\ \hline 

\end{tabular}
\end{center}
\caption{Parameters of the CSS-T codes in \cite{csst-RM,csst-us}, and the codes given in this section.}\label{table:jcss-t}
\end{table}

\section{Private Information Retrieval}\label{sec:PIR}

A \emph{Private Information Retrieval} (PIR) scheme is a cryptographic protocol that enables a user to retrieve an item from a database without revealing to the database owner which item is being accessed. When the data is stored across multiple servers, that is, in a distributed storage system, no individual server can determine the specific item requested by the user. In this latter case, one proposed approach for constructing PIR schemes involves encoding the data using a storage linear code $C \subseteq \mathbb{F}_q^n$, and employing a retrieval linear code $D \subseteq \mathbb{F}_q^n$ for the data retrieval process~\cite{PIR-RM}. 

 If any set of $t$ servers cannot obtain any information about the requested item, the PIR scheme is said to resist a $t$-collusion attack, or equivalently, to provide privacy level $t$.
The following result characterizes the security and efficiency of a PIR scheme under collusion attacks, where some servers may share their data in an attempt to infer the user's request  based on such a pair $C, D$ of linear codes. Note that a linear code is said to be \emph{transitive} if its \emph{automorphism group}  acts transitively on the set of coordinates. That is, for any pair of coordinate positions $i, j \in \{1, \dots, n\}$, there exists a permutation $\pi$ of the coordinate positions such that $\pi(i) = j$ and $\pi$ is an automorphism of the code $C$. We will denote the automorphism group of $C$ by $\operatorname{Aut}(C)$.

\begin{theorem}[\cite{PIR-RM}]\label{th:PIR}
If $\operatorname{Aut}(C)$ and $\operatorname{Aut}(C\star D)$ act transitively on the set of coordinates $\{1,\dots,n\}$, then there exists a PIR scheme with rate $$R=\frac{\dim(C\star D)^{\bot}}{n}$$ such that it resists a $(d({D^{\perp}})-1)$--collusion attack.\end{theorem}

In the framework of binary codes or codes over a small field relative to their length,  Reed--Muller codes~\cite{PIR-RM} and cyclic codes~\cite{PIRcyclic} have been successfully proposed to address the properties outlined in the theorem above and to construct PIR schemes. This is also the framework for this work and we will show that PIR schemes based on hyperbolic codes may outperform those constructed from Reed--Muller codes over non-binary fields. Moreover, we will consider $J$-variety codes and their subfield subcodes to construct PIR schemes with excellent parameters.

\subsection{Transitivity}

In order to propose new families of codes for constructing PIR schemes  we must ensure that the codes used are transitive. In~\cite{decreasing}, a new family of monomial-Cartesian codes, known as decreasing monomial-Cartesian codes, was introduced. These codes are defined by evaluating monomials in a manner analogous to the construction of Reed--Solomon and Reed--Muller codes. However, they impose additional conditions on the functions to be evaluated, specifically in terms of divisibility.

\begin{definition}[Decreasing monomial-Cartesian code]\label{de:decreasing}
A \emph{decreasing monomial set} is a set of monomials $\mathcal{M} \subseteq R$ such that, if $\mathbf{m} \in \mathcal{M}$ and $\mathbf{m}^\prime$ divides $\mathbf{m}$, then $\mathbf{m}^\prime \in \mathcal{M}$. The code $C_{\Delta,Z}$ is called a \emph{decreasing monomial-Cartesian code} if the set of monomials $\{\mathbf{x}^\mathbf{a} \mid \mathbf{a} \in \Delta \}$ forms a decreasing monomial set.
\end{definition}

In~\cite[Theorem 3.9]{decreasing}, the minimum distance and dimension of a decreasing monomial-Cartesian code are computed using a minimal generating set of $\mathcal{M}$. Additionally, in~\cite[Theorem 3.3]{decreasing}, it is shown that the dual of a decreasing monomial-Cartesian code is equivalent to another decreasing monomial-Cartesian code. The following lemma shows that a decreasing monomial-Cartesian code is transitive.

\begin{lemma}
Let $C_{\Delta,Z}$ be a decreasing monomial-Cartesian code, where $Z$ is an additive subgroup of $\mathbb{F}_q^n$. Then $C_{\Delta,Z}$ is transitive.
\end{lemma}

\begin{proof}
Recall that each coordinate $i \in \{1,\dots,n\}$ can be identified with a point $\mathbf{P}_i \in Z$. Without loss of generality, we assume that a codeword $(c_{\mathbf{P}_1}, \ldots, c_{\mathbf{P}_n})$ is obtained by evaluating a monomial $f(\mathbf{x}) = \mathbf{x}^{\mathbf{a}}$ (a general codeword is simply a linear combination of such evaluations). Given two distinct points $\mathbf{P}_i, \mathbf{P}_j \in Z$, we aim to show that there exists a permutation $\pi: Z \rightarrow Z$ such that $\pi(\mathbf{P}_i) = \mathbf{P}_j$ and that the permuted codeword $(c_{\pi(\mathbf{P}_1)}, \ldots, c_{\pi(\mathbf{P}_n)})$ still lies in $C_{\Delta,Z}$.

Consider the map $\pi(z) = {z} - \mathbf{P}_i + \mathbf{P}_j$. Clearly, $\pi(\mathbf{P}_i) = \mathbf{P}_j$, and $\pi$ defines a permutation on $Z$ since $Z$ is an additive subgroup. Now, define the polynomial
\[
g(\mathbf{x}) = \prod_{s=1}^n \left(x_s - {\mathbf{P}_i}_s + {\mathbf{P}_j}_s\right)^{\mathbf{a}_s}.
\]
Since the monomial set defining the code is decreasing, $g(\mathbf{x})$ is a linear combination of monomials in $\Delta$, as each such monomial divides $f(\mathbf{x}) = \mathbf{x}^{\mathbf{a}}$. Therefore, the permuted codeword satisfies
\[
(c_{\pi(\mathbf{P}_1)}, \ldots, c_{\pi(\mathbf{P}_n)}) = (f \circ \pi(\mathbf{P}_1), \ldots, f \circ \pi(\mathbf{P}_n)) = (g(\mathbf{P}_1), \ldots, g(\mathbf{P}_n)) \in C_{\Delta,Z}.
\]
\end{proof}

In the context of $J$-affine codes, being a decreasing monomial-Cartesian code follows from considering \emph{consecutive} cyclotomic sets and it is widely used (see, for example,~\cite{Jaffine15,Jaffine17}). That is, $\Delta_{\boldsymbol{a}_i}= I_{\boldsymbol{a}_0} \cup I_{\boldsymbol{a}_1} \cup \cdots \cup I_{\boldsymbol{a}_i}$.

\begin{lemma}\label{lem:jafin-trans}
Let $C^J_\Delta$ be a $J$-affine code defined by a union of consecutive cyclotomic sets $\Delta = \Delta_{\boldsymbol{a}_i}= I_{\boldsymbol{a}_0} \cup I_{\boldsymbol{a}_1} \cup \cdots \cup I_{\boldsymbol{a}_i}$. Then $\Delta$ is a decreasing monomial set and  $C^J_\Delta$ is transitive.
\end{lemma}

\begin{proof}
The proof follows similar reasoning to the previous lemma, taking into account that
\[
g(\mathbf{x}) = \prod_{s=1}^n \left(x_s - {\mathbf{P}_i}_s + {\mathbf{P}_j}_s\right)^{\mathbf{a}_s \cdot q^k} = \prod_{s=1}^n \left(x_s - {\mathbf{P}_i}_s + {\mathbf{P}_j}_s\right)^{\mathbf{a}_s},
\]
since the exponents are taken modulo the size of the field.
\end{proof}

\begin{lemma}
Let $C_{\Delta_1,Z}$ and $C_{\Delta_2,Z}$ be two transitive monomial-Cartesian codes. Then the componentwise (or star) product code $C_{\Delta_1,Z} \star C_{\Delta_2,Z}$ is also transitive.
\end{lemma}

\begin{proof}
This follows directly from the fact, noted in Section~\ref{se:dos}, that the $\star$-product code corresponds to the evaluation code defined by the sum of the monomial sets $\Delta_1 + \Delta_2$.
\end{proof}

\begin{remark}
Note that a code and its dual share the same automorphism group, see for example~\cite{Huffman_Pless_2003}. Therefore, if a code is transitive, so it is its dual. This observation, together with the preceding lemmas, allows us to consider decreasing monomial-Cartesian codes, hyperbolic codes, and their duals for the construction of PIR schemes in coding theory  framework of~\cite{PIR-RM} that provide the parameters in Theorem~\ref{th:PIR}.
\end{remark}

\subsection{PIR from hyperbolic codes}

We begin by addressing the case in which the storage code $C$ is a Reed--Muller code and the retrieval code $D$ is taken to be the dual of a hyperbolic code. We compare this with the classical setting where both $C$ and $D$ are Reed--Muller codes, as in~\cite{PIR-RM}. Specifically, in our comparison, the storage code $C$ is fixed as a Reed--Muller code in both scenarios, while the retrieval code $D$ is either a Reed--Muller code (as in~\cite{PIR-RM}) or the dual of a hyperbolic code with the same minimum distance.

It is important to note that the dimension of a hyperbolic code is greater than or equal to that of a Reed--Muller code with the same minimum distance. Furthermore, since the dual of a Reed--Muller code is again a Reed--Muller code, it follows that the dual of a hyperbolic code has dimension less than or equal to that of a Reed--Muller code with the same minimum distance. This implies that both retrieval codes offer the same level of privacy, resisting a $t$-collusion attack with $t = d(D^\perp) - 1$, as both codes have the same minimum distance. However, if $\dim(\operatorname{Hyp}_q(s,m)) > \dim(\operatorname{RM}_q(s,m))$, the PIR scheme achieves a better rate when $D$ is the dual of a hyperbolic code, since the dimension of $(C \star D)^\perp$ is smaller in this case.

We illustrate the aforementioned behavior over the finite field with $7$ elements using two and three variables. In Table~\ref{table:I} and Table~\ref{table:II}, we fix the storage code $C$ as the Reed--Muller codes $\operatorname{RM}_{7}(1,2)$ and $\operatorname{RM}_{7}(1,3)$, respectively. For the retrieval code $D$, the shaded rows correspond to the case where $D$ is a Reed--Muller code, while the bold rows correspond to the case where $D$ is the dual of a hyperbolic code, both with the same minimum distance. We observe that the duals of hyperbolic codes consistently yield better rates than their Reed--Muller counterparts.

\begin{table}[h]
\begin{center}
\begin{tabular}{|c|c|c|c|c|c|c|c|}
\hline
s & $ C $  & $D$ & $D^{\bot}$ & $C\star D$ & ${(C\star D)}^\bot$ & Privacy & $R_{PIR}$  \\\hline \rowcolor{Gray}
3 & $[49,3,42]_7$  & $[49,10,28]_7$&$[49,39,5]_7$ & $[49,15,21]_7$ & $[49,34,6]_7$ & 4 & $34/49$ \\ \hline
\textbf{5} & $\textbf{[49,3,42]}_7$  & $\textbf{[49,8,28]}_7$&$\textbf{[49,41,5]}_7$ & $\textbf{[49,14,21]}_7$ & $\textbf{[49,35,6]}_7$ & \textbf{4} & $\textbf{35/49}$ \\ \hline
\rowcolor{Gray}
4 &  $[49,3,42]_7$  & $[49,15,21]_7$&$[49,34,6]_7$ & $[49,21,14]_7$ & $[49,28,7]_7$ & 5 & $28/49$ \\ \hline
\textbf{6} & $\textbf{[49,3,42]}_7$  & $\textbf{[49,10,21]}_7$&$\textbf{[49,39,6]}_7$ & $\textbf{[49,18,14]}_7$ & $\textbf{[49,31,7]}_7$ & \textbf{5} & $\textbf{31/49}$ \\ \hline
\rowcolor{Gray}
5 &  $[49,3,42]_7$  & $[49,21,14]_7$&$[49,28,7]_7$ & $[49,28,7]_7$ & $[49,21,14]_7$ & 6 & $21/49$ \\ \hline
\textbf{7} &  $\textbf{[49,3,42]}_7$  & $\textbf{[49,14,14]}_7$&$\textbf{[49,35,7]}_7$ & $\textbf{[49,23,7]}_7$ & $\textbf{[49,26,12]}_7$ & \textbf{6} & $\textbf{26/49}$ \\ \hline
\rowcolor{Gray}
6 & $[49,3,42]_7$  & $[49,28,7]_7$&$[49,21,14]_7$ & $[49,34,6]_7$ & $[49,15,21]_7$ & 13 & $15/49$ \\ \hline
\textbf{14} & $\textbf{[49,3,42]}_7$  & $\textbf{[49,25,7]}_7$&$\textbf{[49,24,14]}_7$ & $\textbf{[49,32,6]}_7$ & $\textbf{[49,17,20]}_7$ & \textbf{13} & $\textbf{17/49}$ \\ \hline
\rowcolor{Gray}
7 & $[49,3,42]_7$  & $[49,34,6]_7$&$[49,15,21]_7$ & $[49,39,5]_7$ & $[49,10,28]_7$ & 20 & $10/49$ \\ \hline
\textbf{21} & $\textbf{[49,3,42]}_7$  & $\textbf{[49,34,6]}_7$&$\textbf{[49,15,21]}_7$ & $\textbf{[49,39,5]}_7$ & $\textbf{[49,10,28]}_7$ & \textbf{20} & $\textbf{10/49}$ \\ \hline

\end{tabular}
\caption{Comparison of $D = \operatorname{RM}_7(s,2)$ codes (shaded rows) with $D = \operatorname{Hyp}_{7}(s,2)^\perp$ codes (boldface rows).}\label{table:I}
\end{center}
\end{table} 

\begin{table}[h]
\begin{center}
\begin{tabular}{|c|c|c|c|c|c|c|c|}
\hline
s & $ C $  & $D$ & $D^{\bot}$ & $C\star D$ & ${(C\star D)}^\bot$ & Privacy & $R_{PIR}$  \\\hline \rowcolor{Gray}
2 & $[343,4,294]_7$  & $[343,10,245]_7$&$[343,333,4]_7$ & $[343,20]_7$ & $[343,323]_7$ & 3 & $323/343$ \\ \hline
\textbf{4} & $\textbf{[343,4,294]}_7$  & $\textbf{[343,7,245]}_7$&$\textbf{[343,336,4]}_7$ & $\textbf{[343,19]}_7$ & $\textbf{[343,324]}_7$ & \textbf{3} & $\textbf{324/343}$ \\ \hline
\rowcolor{Gray}
3 & $[343,4,294]_7$  & $[343,20,196]_7$&$[343,323,5]_7$ & $[343,35]_7$ & $[343,308]_7$ & 4 & $308/343$ \\ \hline
\textbf{5} & $\textbf{[343,4,294]}_7$  & $\textbf{[343,13,21]}_7$&$\textbf{[343,330,5]}_7$ & $\textbf{[343,29]}_7$ & $\textbf{[343,314]}_7$ & \textbf{4} & $\textbf{314/343}$ \\ \hline
\rowcolor{Gray}
4 & $[343,4,294]_7$  & $[343,35,147]_7$&$[343,308,6]_7$ & $[343,56]_7$ & $[343,287]_7$ & 5 & $287/343$ \\ \hline
\textbf{6} & $\textbf{[343,4,294]}_7$  & $\textbf{[343,16,147]}_7$&$\textbf{[343,327,6]}_7$ & $\textbf{[343,38]}_7$ & $\textbf{[343,305]}_7$ & \textbf{5} & $\textbf{305/343}$ \\ \hline
\rowcolor{Gray}
5 & $[343,4,294]_7$  & $[343,56,98]_7$&$[343,287,7]_7$ & $[343,84]_7$ & $[343,259]_7$ & 6 & $259/343$ \\ \hline
\textbf{7} & $\textbf{[343,4,294]}_7$  & $\textbf{[343,25,98]}_7$&$\textbf{[343,318,7]}_7$ & $\textbf{[343,53]}_7$ & $\textbf{[343,290]}_7$ & \textbf{6} & $\textbf{290/343}$ \\ \hline
\rowcolor{Gray}
6 & $[343,4,294]_7$  & $[343,84,49]_7$&$[343,259,14]_7$ & $[343,117]_7$ & $[343,226]_7$ & 13 & $226/343$ \\ \hline
\textbf{14} & $\textbf{[343,4,294]}_7$  & $\textbf{[343,59,49]}_7$&$\textbf{[343,284,14]}_7$ & $\textbf{[343,98]}_7$ & $\textbf{[343,245]}_7$ & \textbf{13} & $\textbf{245/343}$ \\ \hline
\rowcolor{Gray}
7 & $[343,4,294]_7$  & $[343,117,42]_7$&$[343,226,21]_7$ & $[343,153]_7$ & $[343,190]_7$ & 20 & $190/343$ \\ \hline
\textbf{21} & $\textbf{[343,4,294]}_7$  & $\textbf{[343,95,42]}_7$&$\textbf{[343,248,21]}_7$ & $\textbf{[343,144]}_7$ & $\textbf{[343,199]}_7$ & \textbf{20} & $\textbf{199/343}$ \\ \hline
\rowcolor{Gray}
8 & $[343,4,294]_7$  & $[343,153,35]_7$&$[343,190,28]_7$ & $[343,190]_7$ & $[343,153]_7$ & 27 & $153/343$ \\ \hline
\textbf{28} & $\textbf{[343,4,294]}_7$  & $\textbf{[343,120,35]}_7$&$\textbf{[343,223,28]}_7$ & $\textbf{[343,154]}_7$ & $\textbf{[343,169]}_7$ & \textbf{27} & $\textbf{169/343}$ \\ \hline
\rowcolor{Gray}
9 & $[343,4,294]_7$  & $[343,190,28]_7$&$[343,153,35]_7$ & $[343,226]_7$ & $[343,117]_7$ & 34 & $117/343$ \\ \hline
\textbf{35} & $\textbf{[343,4,294]}_7$  & $\textbf{[343,144,28]}_7$&$\textbf{[343,199,35]}_7$ & $\textbf{[343,201]}_7$ & $\textbf{[343,142]}_7$ & \textbf{34} & $\textbf{142/343}$ \\ \hline
\rowcolor{Gray}
10 & $[343,4,294]_7$  & $[343,226,21]_7$&$[343,117,42]_7$ & $[343,259]_7$ & $[343,84]_7$ & 41 & $84/343$ \\ \hline
\textbf{42} & $\textbf{[343,4,294]}_7$  & $\textbf{[343,168,21]}_7$&$\textbf{[343,175,42]}_7$ & $\textbf{[343,225]}_7$ & $\textbf{[343,118]}_7$ & \textbf{41} & $\textbf{118/343}$ \\ \hline
\rowcolor{Gray}
11 & $[343,4,294]_7$  & $[343,259,14]_7$&$[343,84,49]_7$ & $[343,287]_7$ & $[343,56]_7$ & 48 & $56/343$ \\ \hline
\textbf{49} & $\textbf{[343,4,294]}_7$  & $\textbf{[343,192,14]}_7$&$\textbf{[343,151,49]}_7$ & $\textbf{[343,244]}_7$ & $\textbf{[343,99]}_7$ & \textbf{48} & $\textbf{99/343}$ \\ \hline
\rowcolor{Gray}
12 & $[343,4,294]_7$  & $[343,287,7]_7$&$[343,56,98]_7$ & $[343,308]_7$ & $[343,35]_7$ & 97 & $35/343$ \\ \hline
\textbf{98} & $\textbf{[343,4,294]}_7$  & $\textbf{[343,265,7]}_7$&$\textbf{[343,78,98]}_7$ & $\textbf{[343,295]}_7$ & $\textbf{[343,48]}_7$ & \textbf{97} & $\textbf{48/343}$ \\ \hline
\end{tabular}
\end{center}
\caption{Comparison of $D = \operatorname{RM}_7(s,3)$ codes (shaded rows) with $D = \operatorname{Hyp}_7(s,3)^\perp$ codes (boldface rows).}
\label{table:II}
\end{table}

\subsection{PIR with subfield-subcodes of $J$-affine variety codes}

We propose two constructions that provide pairs of codes with the desired properties to construct a PIR scheme from subfield subcodes of $J$-affine variety codes.

\subsubsection{Using subfield subcodes of $J$-Affine Variety Codes in One Variable}

It is important to note that subfield subcodes of $J$-affine variety codes correspond to BCH codes when the evaluation does not include zero; however, this correspondence no longer holds when evaluation at zero is considered. Let \( q^r \) denote the order of the finite field, and \( q \) the order of the subfield considered, that is, the cyclotomic cosets are computed over \( \mathbb{F}_q \). Let \( n \) be a divisor of \( q^r - 1 \), and let \( I_a \) denote the cyclotomic coset associated with \( a \) modulo \( n \). In this setting, the code is the evaluation code at the \( n \)-th roots of unity, possibly including zero.

The subfield subcodes of one-variable $J$-affine codes may lead to new PIR schemes with improved performance. The following example presents some representative parameters that demonstrate this. Using the method described in Section~\ref{sec:PIR}, where $C$ is taken as the code $\operatorname{RM}_7(1,2)$ and $D$ is chosen as the dual of a hyperbolic code, the parameters obtained by puncturing these codes are shown in the shaded rows of Table~\ref{tab:cyclic}.

It should be noted that puncturing or shortening a transitive code is a valid form of comparison, as it preserves the minimum distance of the original code (see~\cite[Theorem~7.6.1]{Huffman_Pless_2003}), and results in a code of similar length to the subfield subcode being compared. Furthermore, the transitivity of the code ensures that the result of puncturing is independent of the specific positions chosen, since all punctured versions are permutation equivalent (see~\cite[Theorem~1.6.6]{Huffman_Pless_2003}).

\begin{remark}
To simplify the notation, in the remainder of the paper, we will denote the subfield subcode of the $J$-affine variety codes $C^{J,\sigma}_{\Delta_C}$ and $D^{J,\sigma}_{\Delta_D}$, by just $C$ and $D$, respectively. We will also say that $C$ and $D$ are defined by $\Delta_C$ and $\Delta_D$, respectively, and that $C$ and $D$ have defining set $\Delta_C$ and $\Delta_D$, respectively.
\end{remark}

\begin{example}
Consider \( q = 7 \), \( r = 2 \), \( N_1 = 48 \), and set \( J = \{1\} \). Then \( \mathcal{R}_J = \mathbb{F}_q[x_1]/\langle x_1^{48} - 1 \rangle \). Let \( \Delta_C = \{24, 25, 31\} \) and \( \Delta_D = \{24, 25, 31, 32\} \). The code \( C \) is a \([48,3]_7\) code and \( D \) is a \([48,4]_7\) code, whose dual \( D^\perp \) has parameters \([48,44,4]_7\). Therefore, the Schur product code \( C \star D \) has parameters \([48,8]_7\), and its dual \( (C \star D)^\perp \) has parameters \([48,40]_7\).

Since both \( C \) and \( D \) are cyclic codes, and the automorphism group of a cyclic code is transitive (see~\cite[Theorem~1.6.4]{Huffman_Pless_2003}), Theorem~\ref{th:PIR} is applicable. Consequently, this PIR scheme is secure against 3 colluding servers and achieves a PIR rate of \(  \frac{40}{48} \).

At the same privacy level, the punctured code \( D \), which is the dual of a hyperbolic code—specifically, \( D^\perp \) is the shortened hyperbolic code \( \operatorname{Hyp}_7(4,2) \)—has parameters \([48,5]_7\) and provides a PIR rate of \(  \frac{38}{48} \). Hence, the subfield subcode of the $J$-affine code achieves a better PIR rate than a hyperbolic code. These parameters are shown in the first two rows of Table~\ref{tab:cyclic}. The first row, shaded in gray, corresponds to the code generated using the method described in Section~\ref{sec:PIR}, while the bold-faced row represents the code obtained as a subfield subcode of a one-variable $J$-affine variety code.

In Table~\ref{tab:cyclic}, except for the case corresponding to privacy level 23, all subfield subcodes of one-variable $J$-affine codes either achieve the same PIR rate, or outperform those based on hyperbolic duals for the same privacy level. Moreover, the use of cyclotomic cosets allows a larger set of design parameters, enabling the construction of PIR schemes with increased privacy levels and improved rates. In some cases, suitable parameter sets for comparison may not exist; however, the obtained parameters still contribute meaningfully to get a broader set of achievable PIR parameters configurations.

\begin{table}[h]
\centering % centering table
\begin{tabular}{|c|c |c| c| c| c| c|c |} % creating 10 columns
\hline % inserting double-line
$s$ & $C $  & $D$ & $D^{\bot}$ & $C\star D$ & ${(C\star D)}^\bot$  & Privacy & $R_{PIR}$ 
\\ \rowcolor{Gray}
\hline % inserts single-line
$4$&$[48,3]_{7}$&$[48,5]_{7}$ &$[48,43,4]_{7}$&$[48,10]_{7}$ & $[48,38]_7$ & $3$ & $38/48$ \\
 \hline
&$\mathbf{[48,3]_{7}}$&$\mathbf{[48,4]_{7}}$ &$\mathbf{[48,44,4]_{7}}$&$\mathbf{[48,8]_{7}}$ & $\mathbf{[48,40]_7}$ & $\mathbf{3}$ & $\mathbf{40/48}$ \\
 \hline\rowcolor{Gray}
$5$& $[48,3]_{7}$&$[48,8]_{7}$ &$[48,40,5]_{7}$&$[48,14]_{7}$ & $[48,34]_7$ & $4$ & $34/48$ \\
 \hline 
&$\mathbf{[48,3]_{7}}$&$\mathbf{[48,7]_{7}}$ &$\mathbf{[48,41,5]_{7}}$&$\mathbf{[48,14]_{7}}$ & $\mathbf{[48,34]_7}$ & $\mathbf{4}$ & $\mathbf{34/48}$ \\
 \hline\rowcolor{Gray}
$6$&$[48,3]_{7}$&$[48,10]_{7}$ &$[48,38,6]_{7}$&$[48,18]_{7}$ & $[48,30]_7$  &  $5$ & $30/48$  \\
  \hline
&$\mathbf{[48,3]_{7}}$&$\mathbf{[48,9]_{7}}$ &$\mathbf{[48,39,6]_{7}}$&$\mathbf{[48,15]_{7}}$ & $\mathbf{[48,33]_7}$  &  $\mathbf{5}$ & $\mathbf{33/48}$  \\
  \hline \rowcolor{Gray}
 $8$& $[48,3]_{7}$&$[48,16]_{7}$ &$[48,32,8]_{7}$&$[48,25]_{7}$ & $[48,23]_7$  & $7$ & $23/48$  \\
\hline % inserts single-line
&$\mathbf{[48,3]_{7}}$&$\mathbf{[48,13]_{7}}$ &$\mathbf{[48,35,8]_{7}}$&$\mathbf{[48,23]_{7}}$ & $\mathbf{[48,25]_7}$  & $\mathbf{7}$ & $\mathbf{25/48}$  \\
\hline  \rowcolor{Gray}
$9$&$[48,3]_{7}$&$[48,18]_{7}$ &$[48,30,9]_{7}$&$[48,27]_{7}$ & $[48,21]_7$  & $8$ & $21/48$  \\
\hline % inserts single-line
&$\mathbf{[48,3]_{7}}$&$\mathbf{[48,16]_{7}}$ &$\mathbf{[48,32,9]_{7}}$&$\mathbf{[48,26]_{7}}$ & $\mathbf{[48,22]_7}$  & $\mathbf{8}$ & $\mathbf{21/48}$  \\
\hline  \rowcolor{Gray}
$12$&$[48,3]_{7}$&$[48,21]_{7}$ &$[48,27,12]_{7}$&$[48,29]_{7}$ & $[48,19]_7$  & $11$ & $19/48$  \\
\hline % inserts single-line
&$\mathbf{[48,3]_{7}}$&$\mathbf{[48,18]_{7}}$ &$\mathbf{[48,30,12]_{7}}$&$\mathbf{[48,28]_{7}}$ & $\mathbf{[48,19]_7}$  & $\mathbf{11}$ & $\mathbf{19/48}$  \\
\hline % inserts single-line
&$\mathbf{[48,3]_{7}}$&$\mathbf{[48,20]_{7}}$ &$\mathbf{[48,28,13]_{7}}$&$\mathbf{[48,29]_{7}}$ & $\mathbf{[48,18]_7}$ & $\mathbf{12}$ & $\mathbf{18/48}$ \\
\hline  \rowcolor{Gray}
$14$&$[48,3]_{7}$&$[48,25]_{7}$ &$[48,23,14]_{7}$&$[48,32]_{7}$ & $[48,16]_7$  &$13$ & $16/48$ \\
\hline % inserts single-line
&$\mathbf{[48,3]_{7}}$&$\mathbf{[48,22]_{7}}$ &$\mathbf{[48,26,14]_{7}}$&$\mathbf{[48,31]_{7}}$ & $\mathbf{[48,17]_7}$  &$\mathbf{13}$ & $\mathbf{17/48}$ \\
\hline % inserts single-line
&$\mathbf{[48,3]_{7}}$&$\mathbf{[48,27]_{7}}$ &$\mathbf{[48,21,19]_{7}}$&$\mathbf{[48,34]_{7}}$ & $\mathbf{[48,14]_7}$  & $\mathbf{18}$& $\mathbf{14/48}$ \\
\hline  \rowcolor{Gray}
$20$&$[48,3]_{7}$&$[48,32]_{7}$ &$[48,16,20]_{7}$&$[48,38]_{7}$ & $[48,10]_7$  & $19$& $10/48$ \\
\hline % inserts single-line
&$\mathbf{[48,3]_{7}}$&$\mathbf{[48,29]_{7}}$ &$\mathbf{[48,19,20]_{7}}$&$\mathbf{[48,36]_{7}}$ & $\mathbf{[48,12]_7}$  & $\mathbf{19}$& $\mathbf{12/48}$ \\
\hline  \rowcolor{Gray}
$21$&$[48,3]_{7}$&$[48,34]_{7}$ &$[48,14,21]_{7}$&$[48,39]_{7}$ & $[48,9]_7$  &$20$ & $9/48$ \\
\hline % inserts single-line
&$\mathbf{[48,3]_{7}}$&$\mathbf{[48,31]_{7}}$ &$\mathbf{[48,17,21]_{7}}$&$\mathbf{[48,38]_{7}}$ & $\mathbf{[48,10]_7}$  &$\mathbf{20}$ & $\mathbf{10/48}$ \\
\hline % inserts single-line
&$\mathbf{[48,3]_{7}}$&$\mathbf{[48,33]_{7}}$ &$\mathbf{[48,15,22]_{7}}$&$\mathbf{[48,40]_{7}}$ & $\mathbf{[48,8]_7}$  &$\mathbf{21}$ &$\mathbf{8/48}$  \\
\hline \rowcolor{Gray}
$24$&$[48,3]_{7}$&$[48,36]_{7}$ &$[48,12,24]_{7}$&$[48,41]_{7}$ & $[48,7]_7$ & $23$& $7/48$  \\
\hline % inserts single-line
&$\mathbf{[48,3]_{7}}$&$\mathbf{[48,35]_{7}}$ &$\mathbf{[48,13,24]_{7}}$&$\mathbf{[48,42]_{7}}$ & $\mathbf{[48,6]_7}$ & $\mathbf{23}$& $\mathbf{6/48}$  \\
\hline % inserts single-line
&$\mathbf{[48,3]_{7}}$&$\mathbf{[48,40]_{7}}$ &$\mathbf{[48,8,33]_{7}}$&$\mathbf{[48,44]_{7}}$ & $\mathbf{[48,4]_7}$ &$\mathbf{32}$ &$\mathbf{4/48}$   \\
\hline % inserts single-line
&$\mathbf{[48,3]_{7}}$&$\mathbf{[48,42]_{7}}$ &$\mathbf{[48,6,34]_{7}}$&$\mathbf{[48,45]_{7}}$ & $\mathbf{[48,3]_7}$  &$\mathbf{33}$ & $\mathbf{3/48}$ \\\hline\rowcolor{Gray}
$35$&$[48,3]_{7}$&$[48,43]_{7}$ &$[48,5,35]_{7}$&$[48,46]_{7}$ & $[48,2]_7$  &$34$ & $2/48$ \\\hline
&$\mathbf{[48,3]_{7}}$&$\mathbf{[48,43]_{7}}$ &$\mathbf{[48,5,35]_{7}}$&$\mathbf{[48,46]_{7}}$ & $\mathbf{[48,2]_7}$  &$\mathbf{34}$ & $\mathbf{2/48}$ \\\hline
\end{tabular}
\caption{Comparison of shortened \( D = \operatorname{Hyp}_7(s,2)^\perp \) code (shaded rows) with subfield subcode of \( J \)-affine code (boldface rows)} \label{tab:cyclic}
\end{table}
\begin{table}[h]
\begin{center}
\begin{tabular}{l|l}
\hline
$ \Delta_1 $  &  $\Delta_2$    \\ \hline
 $ I_{24} \cup I_{25}$ &  $ I_{24} \cup I_{25} \cup I_{32} $   \\
 $ $ &  $  I_{25} \cup I_{32}\cup I_{33} \cup I_{34} $   \\
$   $ &   $ I_{24} \cup I_{25} \cup I_{32}\cup I_{33} \cup I_{34}\cup  I_{40}$\\
$   $ &   $ I_{24} \cup I_{25} \cup I_{32}\cup I_{33} \cup I_{34}\cup  I_{40}\cup  I_{5}\cup  I_{18}$ \\
$   $ &  $  I_{25} \cup I_{32}\cup I_{33} \cup I_{34}\cup  I_{40}\cup  I_{5}\cup  I_{18}\cup  I_{12}\cup  I_{19}$  \\
$   $ &  $  I_{25} \cup I_{32}\cup I_{33} \cup I_{34}\cup  I_{40}\cup  I_{5}\cup  I_{18}\cup  I_{12}\cup  I_{19}\cup  I_{26} $ \\
$   $ &  $ I_{25} \cup I_{32}\cup I_{33} \cup I_{34}\cup  I_{40}\cup  I_{5}\cup  I_{18}\cup  I_{12}\cup  I_{19}\cup  I_{26} \cup I_{41} $ \\
$   $ &   $ I_{25} \cup I_{32}\cup I_{33} \cup I_{34}\cup  I_{40}\cup  I_{5}\cup  I_{18}\cup  I_{12}\cup  I_{19}\cup  I_{26} \cup I_{41}\cup I_{11} $ \\
$   $ &   $ I_{24} \cup I_{25} \cup I_{32}\cup I_{33} \cup I_{34}\cup  I_{40}\cup  I_{5}\cup  I_{18}\cup  I_{12}\cup  I_{19}\cup  I_{26} \cup I_{41}\cup I_{11}\cup I_{4}\cup I_{27} $  \\
$   $ &   $ I_{24} \cup I_{25} \cup I_{32}\cup I_{33} \cup I_{34}\cup  I_{40}\cup  I_{5}\cup  I_{18}\cup  I_{12}\cup  I_{19}\cup  I_{26} \cup I_{41}\cup I_{11}\cup I_{4}\cup I_{27}\cup I_{6} $  \\
$   $ &   $ I_{24} \cup I_{25} \cup I_{32}\cup I_{33} \cup I_{34}\cup  I_{40}\cup  I_{5}\cup  I_{18}\cup  I_{12}\cup  I_{19}\cup  I_{26} \cup I_{41}\cup I_{11}\cup I_{4}\cup I_{27}\cup I_{6}\cup I_{17} $  \\
$   $ &   $ I_{24} \cup I_{25} \cup I_{32}\cup I_{33} \cup I_{34}\cup  I_{40}\cup  I_{5}\cup  I_{18}\cup  I_{12}\cup  I_{19}\cup  I_{26} \cup I_{41}\cup I_{11}\cup I_{4}\cup I_{27}\cup I_{6}\cup I_{17}\cup I_{10} $  \\
$   $ &   $ I_{24} \cup I_{25} \cup I_{32}\cup I_{33} \cup I_{34}\cup  I_{40}\cup  I_{5}\cup  I_{18}\cup  I_{12}\cup  I_{19}\cup  I_{26} \cup I_{41}\cup I_{11}\cup I_{4}\cup I_{27}\cup I_{6}\cup I_{17}\cup I_{10}\cup I_{13} $  \\
$   $ &   $ I_{24} \cup I_{25} \cup I_{32}\cup I_{33} \cup I_{34}\cup  I_{40}\cup  I_{5}\cup  I_{18}\cup  I_{12}\cup  I_{19}\cup  I_{26} \cup I_{41}\cup I_{11}\cup I_{4}\cup I_{27}\cup I_{6}\cup I_{17}\cup I_{10}\cup I_{13}\cup I_{20}\cup I_{0}\cup I_{3} $  \\
$   $ &   $ I_{24} \cup I_{25} \cup I_{32}\cup I_{33} \cup I_{34}\cup  I_{40}\cup  I_{5}\cup  I_{18}\cup  I_{12}\cup  I_{19}\cup  I_{26} \cup I_{41}\cup I_{11}\cup I_{4}\cup I_{27}\cup I_{6}\cup I_{17}\cup I_{10}\cup I_{13}\cup I_{20}\cup I_{0}\cup I_{3}\cup I_{1} $  \\
$   $ &   $ I_{24} \cup I_{25} \cup I_{32}\cup I_{33} \cup I_{34}\cup  I_{40}\cup  I_{5}\cup  I_{18}\cup  I_{12}\cup  I_{19}\cup  I_{26} \cup I_{41}\cup I_{11}\cup I_{4}\cup I_{27}\cup I_{6}\cup I_{17}\cup I_{10}\cup I_{13}\cup I_{20}\cup I_{0}\cup I_{3}\cup I_{1} \cup I_{16}$  \\
\end{tabular}
\end{center}
\caption{Cyclotomic cosets used in constructing the boldface rows in Table \ref{tab:cyclic}. }\label{table:cyclic.cosets}
\end{table}
\end{example}

For dealing with only one variable, we will introduce an alternative approach. Let \( \Delta_D \) be defined as the union of consecutive cyclotomic cosets, specifically \( I_0 \cup I_1 \cup \cdots \cup I_{a_i} \). By the definition of $\mathcal{A} = \{a_0 < a_1 < \cdots \}$, the set of ordered representatives of the minimal cyclotomic cosets (see section \ref{se:subf}), we have that all the elements  smaller than $a_{i+1}$ belong to $I_0 \cup \cdots \cup I_{a_i}$, since  $a_{i+1}$ is the representative of its minimal cyclotomic coset. Thus, we know that the set $$  \{0, 1, 2, \dots, a_{i+1} - 1\} \subseteq \Delta_D, $$ which implies that the minimum distance satisfies \( d(D^\perp) \geq a_{i+1} + 1 \) (BCH bound).
Assume that \( N \) is a divisor of \( q^r - 1 \), and define \( \nu = \frac{q^r - 1}{N} \). We propose two possible definitions for the set \( \Delta_C \):

\begin{enumerate}
    \item \( \Delta_C = \{0, N, 2N, \dots, (\nu - 1)N\} \). Since \( \Delta_1 \) is a union of cyclotomic cosets, the evaluation at these points increases the dimension of the subfield subcode by \( \#\Delta_C \) units.
    \item \( \Delta_C = I_0 \cup I_N \). This is also a union of cyclotomic cosets and contributes to the dimension of the subfield subcode in the same manner as in the previous case.
\end{enumerate}

Note that in some cases, both definitions coincide. Based on their definitions, we have that

\[
\#(\Delta_C + \Delta_D) \leq \#\Delta_C \cdot \#\Delta_D.
\]

Therefore, the dimension of the dual code satisfies the following inequality.

\[
\dim ((C \star D)^\perp) \geq n - \#\Delta_C \cdot \#\Delta_D.
\]

We summarize these ideas in the following result.

\begin{lemma} With the construction given above.
    There exists a PIR scheme with a storage code \( C \) of length \( n \) and dimension $v$ , privacy level \( a_{i+1} \), and rate $$ \frac{n - \#\Delta_C \cdot \#\Delta_D}{n}. $$
\end{lemma}

\begin{example}\label{ex:subfieldsubcodepironevariable}
    Consider the parameters \( q = 2 \), \( r = 8 \), \( n = 255 \), and \( N = 85 \), which means \( \nu = 3 \). We define the sets \( \Delta_D = I_0 \cup I_1 = \{0, 1, 2, 4, 8, 18, 32, 64, 128\} \) and \( \Delta_C = \{0, 85, 170\} \).

 Consequently,  \( D \) is a \( [255, 9]_2 \) code, while its dual code \( D^\perp \) has parameters \( [255, 246, \geq 4]_2 \). The code \( C \) has parameters \( [255, 3]_2 \). Therefore, the code \( C \star D \) has parameters \( [255, 27]_2 \), and its dual \( (C \star D)^\perp \) has parameters \( [255, 228]_2 \).

    This framework provides a PIR scheme of length 255, with a privacy level of 3, and a rate of \( \frac{228}{255} \). This rate is better than the one presented in Table 6 of \cite{PIRcyclic} for the same privacy level, despite having a lower storage rate. Therefore, our approach improves the constellation of possible parameters.

    Table \ref{table:IV} presents several code parameters following this method. We note that in Table \ref{table:IV}, the BCH bound for the retrieval codes (\( D \)) is sharp and matches their minimum distance.
    Moreover, by evaluating at zero, we also obtain a PIR scheme of length 256, privacy level 3, and rate \( \frac{229}{256} \).
\end{example}

\begin{table}[H]
\begin{center}
\begin{tabular}{|c|c|c|c|c|c|c|}
\hline
$ C $  & $D$ & $D^{\bot}$ & $C\star D$ & ${(C\star D)}^\bot$ & Privacy & $R_{PIR}$  \\\hline 
$[255,3,85]_2$  & $[255,9]_2$&$[255,246,4]_2$ & $[255,27]_2$ & $[255,228]_2$ & 3 & $228/255$ \\ \hline
$[255,3,85]_2$  & $[255,17]_2$&$[255,238,6]_2$ & $[255,51]_2$ & $[255,204]_2$ & 5 & $204/255$ \\ \hline
$[255,3,85]_2$  & $[255,25]_2$&$[255,230,8]_2$ & $[255,75]_2$ & $[255,180]_2$ & 7 & $180/255$ \\ \hline
$[255,3,85]_2$  & $[255,33]_2$&$[255,222,10]_2$ & $[255,99]_2$ & $[255,156]_2$ & 9 & $156/255$ \\ \hline
$[255,3,85]_2$  & $[255,49]_2$&$[255,206,14]_2$ & $[255,123]_2$ & $[255,132]_2$ & 13 & $132/255$ \\ \hline
$[255,3,85]_2$  & $[255,57]_2$&$[255,198,16]_2$ & $[255,147]_2$ & $[255,108]_2$ & 15 & $108/255$ \\ \hline
$[255,3,85]_2$  & $[255,65]_2$&$[255,190,18]_2$ & $[255,171]_2$ & $[255,84]_2$ & 17 & $84/255$ \\ \hline
$[255,3,85]_2$  & $[255,69]_2$&$[255,186,20]_2$ & $[255,183]_2$ & $[255,72]_2$ & 19 & $72/255$ \\ \hline
\end{tabular}
\end{center}
\caption{Subfield subcodes of one-variable $J$-affine codes of length $255$ (Example~\ref{ex:subfieldsubcodepironevariable}).}\label{table:IV}
\end{table}
\begin{table}[H]
\begin{center}
\begin{tabular}{l|l}
\hline
$ \Delta_C $  &  $\Delta_D$    \\ \hline
$ \{0, 85, 170\}$ & $I_0 \cup I_1 $   \\
$   $ &  $I_0 \cup I_1\cup I_3  $ \\
$   $ &  $I_0 \cup I_1\cup I_3 \cup I_5 $ \\
$   $ &  $I_0 \cup I_1\cup I_3 \cup I_5 \cup I_7  $  \\
$   $ &   $I_0 \cup I_1\cup I_3 \cup I_5 \cup I_7 \cup I_9 \cup I_{11}  $   \\
$   $ &   $I_0 \cup I_1\cup I_3 \cup I_5 \cup I_7 \cup I_9 \cup I_{11} \cup I_{13}  $   \\
$   $ &   $I_0 \cup I_1\cup I_3 \cup I_5 \cup I_7 \cup I_9 \cup I_{11} \cup I_{13} \cup I_{15}  $   \\
$   $ &   $I_0 \cup I_1\cup I_3 \cup I_5 \cup I_7 \cup I_9 \cup I_{11} \cup I_{13} \cup I_{15}  \cup I_{17}  $   \\
\end{tabular}
\end{center}
\caption{Cyclotomic cosets used for   codes in Table \ref{table:IV}. }\label{table:V}
\end{table}

\begin{lemma}
    If $q=2$ and $r$ is even then $3\mid q^r-1$, therefore there exist a PIR scheme with $q^r-1$ servers, privacy 3, and rate $$\frac{q^r-1-(3(r+1))}{q^r-1}.$$
\end{lemma}

\subsubsection{Using Hyperbolic Codes}

Consider the dual of a hyperbolic code, \( D^{\prime} = \mathrm{Hyp}_q(s, m)^\perp \), with defining set \( \Delta_{D^{\prime}} \). For each point \( P \in \Delta_{D^{\prime}} \), we associate the corresponding cyclotomic coset \( I_P \). We then define the new set \( \Delta_D = \bigcup_{P \in \Delta_{D^{\prime}}} I_P \). Let \( D \) denote the linear code with defining set \( \Delta_D \). It is straightforward that \( d(D^{\perp}) \geq s \).

\begin{example} \label{exp:binary.n49}
Consider the case of two variables with \( m = 2 \) and \( q = 2 \), i.e. \( q^3 = 8 \). We evaluate at all points with nonzero coordinates, i.e., \( A_1 = A_2 = \{1, \alpha, \ldots, \alpha^{q^3-2}\} \), thus \( n = 49 \).

Let \( D^{\prime} \) be an affine variety code defined by the set \( \Delta_{D^{\prime}} = \{(0,0), (1,0), (2,0), (0,1), (0,2)\} \). To ensure that the defining set includes complete cyclotomic cosets, we consider the code \( D \) with defining set
\[
\Delta_D = \{(0,0), (1,0), (2,0), (0,1), (0,2)\} \cup \{(4,0), (0,4)\}.
\]
Clearly, we have \( d(D^{\perp}) \geq 4 \).

Now, define \( C \) as the code with defining set \( \Delta_C = \{(0,0), (1,0), (2,0), (4,0)\} \). Then
\begin{align*}
C \star D = 
&\{(0,0), (1,0), (2,0), (3,0), (4,0), (5,0), (6,0), (0,1), (0,2), (0,4), (1,1), (2,1), (4,1), \\
&(1,2), (2,2), (4,2), (1,4), (2,4), (4,4)\}.
\end{align*}
Therefore, the dimension of \( (C \star D)^{\perp} \) is  \( 30 \).
\end{example}

We now present a specific scenario where the parameters can be explicitly computed, providing PIR schemes with favorable parameters.
Consider affine variety codes in two variables. Let \( N_1 - 1 \mid q^r - 1 \) and \( N_2 - 1 \mid q - 1 \). This choice of \( N_2 \) ensures that the points of the form \( (0,a) \) form minimal cyclotomic cosets, facilitating a selection of the set \( \Delta_1 \) that minimizes the number of elements in the Schur product \( C \star D \).

As in the previous setting, consider the hyperbolic code \( \mathrm{Hyp}_q(s, 2)^\perp \) with defining set \( \Delta_{D} = \bigcup_{P \in \Delta_{D^{\prime}}} I_P \), and define \( D \) as the linear code with this set. It is clear that \( d(D^{\perp}) \geq s \).
Next, define \( C \) as the linear code with defining set
\[
\Delta_C = \{(0,0), (0,1), (0,2), \ldots, (0,a)\}.
\]
Note that \( \Delta_C \) contains \( a+1 \) elements, and since the maximum size is  \( q \), we have \( a \leq q-1 \). Therefore $$\dim (C \star D) \leq   (a + 1)\cdot  \dim (D),$$ and this fact motivates the goal of minimizing this dimension.

Let \( \mathcal{A} = \{a_0 = 0 < a_1 = 1 < a_2 < \cdots < a_{\nu} \} \) denote the ordered set of minimal representatives of the cyclotomic sets.

\begin{theorem} \label{te:PIR_Subfield}
Under the assumptions and notation above, consider \( 0, a_1, a_2 \in \mathcal{A} \), and define the sets:
\[
\Delta_C = I_{(0,0)} \cup I_{(0,1)} \cup I_{(0,2)} = \{(0,0), (0,1), (0,2)\},
\]
\[
\Delta_D = I_{(0,0)} \cup I_{(0,1)} \cup I_{(0,2)} \cup I_{(a_1,0)} \cup I_{(a_2,0)}.
\]
Then, the codes \( C \) and \( D \) defined by $\Delta_C$ and $\Delta_D$ respectively provide a PIR scheme of length \( n_J \), privacy level \( 3 \), and rate at least \( \frac{n_J - (3 \cdot 2r + 5)}{n_J} \).
\end{theorem}

\begin{proof}
From the footprint bound, we find that \( d(D^{\perp}) = 4 \), giving a privacy level 3.

The set \( \Delta_D \) contains two cyclotomic cosets \( I_{(a_1,0)} \) and \( I_{(a_2,0)} \), each of size at most \( r \), and three singleton cosets, implying \( \#\Delta_D \leq 2r + 3 \), and hence \( \dim (D) \leq 2r + 3 \).

To compute \( \dim(C \star D) \), we must consider \( 3 \cdot (2r) \) products between \( \Delta_C \) and the nontrivial cosets in \( \Delta_D \), plus the Minkowski sum \( \Delta_C + \Delta_C \), which includes at most 5 elements if \( q > 4 \). Therefore, \( \#(C \star D) \leq 6r + 5 \), and hence \( \dim ((C \star D)^{\perp}) \geq n_J - (6r + 5) \).
\end{proof}

\begin{example}
Let \( \mathbb{F}_{7^2} \) be the ambient field and \( r = 1 \), that is  the subfield is \( \mathbb{F}_7 \). Take \( N_1 = 49 \), \( N_2 = 7 \), and \( J = \emptyset \), so \( n_J = 343 \). Then, by Theorem~\ref{te:PIR_Subfield}, we obtain a PIR scheme with length \( 343 \), privacy level \( 3 \), and rate \( \frac{326}{343} \). Specifically, the code \( C \) has parameters \([343, 3]_7\), and the dual of \( D \) has parameters \([343, 236, 4]_7\).

Compared to the first two rows in Table~\ref{table:II}, this setup improves the PIR rate. The corresponding rates are \( \frac{323}{343} \) and \( \frac{324}{343} \), respectively, for the same privacy level \( t = 3 \), though our scheme has a lower storage code rate \( R_s = \frac{3}{343} \) versus \( \frac{4}{343} \) in Table~\ref{table:II}.
\end{example}

\begin{proposition} \label{prop:PIR_Subfield_Comparison}
Let \( C \) and \( D \) be subfield subcodes of \( J \)-affine variety codes of lengths \( N_1 \) and \( N_2 \), with \( J = \emptyset \). Assume \( q = 2 \), \( q^r = 2^r \), \( N_1 = 2^r \), and \( N_2 = 2 \). Consider PIR schemes where \( C \) is defined by \( \Delta_C = I_{(0,0)} \), yielding a repetition code \([2^{r+1}, 1]\), i.e., a replicated database. We compare these schemes with those based on Reed--Muller codes under the same privacy level.

\begin{itemize}
\item[\textbf{(a)}] Let
\[
\Delta_D = I_{(0,0)} \cup I_{(1,0)} \cup I_{(0,1)}, \quad \Delta_C = I_{(0,0)}.
\]
Then, the Schur product PIR scheme has length \( n_J = 2^{r+1} \), privacy \( t = 3 \), and rate
\[
\frac{n_J - (r + 2)}{n_J}.
\]
A Reed--Muller scheme with \( C = \mathrm{RM}_2(0, r+1) \) and \( D = \mathrm{RM}_2(1, r+1) \) gives \( D^{\perp} = \mathrm{RM}_2(r-1, r+1) \) with the same privacy and rate.

\item[\textbf{(b)}] Let
\[
\Delta_D = \{ I_{(0,0)}, I_{(0,1)}, I_{(1,1)}, I_{(1,0)}, I_{(3,0)}, I_{(5,0)} \}.
\]
Then, \( \dim(D) \leq 4r + 2 \), and \( d(D^\perp) = 8 \), ensuring privacy \( t = 7 \). The rate is at least:
\[
\frac{n_J - (4r + 2)}{n_J}.
\]

For a Reed--Muller code construction with \( C = \mathrm{RM}_2(0, r+1) \), \( D = \mathrm{RM}_2(2, r+1) \), we also get privacy \( t = 7 \), and
\[
\dim(D) = \binom{r+1}{0} + \binom{r+1}{1} + \binom{r+1}{2}.
\]
Our construction achieves a better PIR rate whenever
\[
\binom{r+1}{0} + \binom{r+1}{1} + \binom{r+1}{2} > 4r + 2, \quad \text{which holds for all } r > 5.
\]
\end{itemize}
\end{proposition}

\begin{example}

We illustrate Proposition~\ref{prop:PIR_Subfield_Comparison}, item~(b), by comparing the subfield subcode construction of $J$-affine variety codes with the Reed--Muller-based PIR scheme under the same level of privacy.

\begin{itemize}

    \item For $r = 7$ and $q = 2$ (i.e. $n_J = 256$), the involved codes have the following parameters.

    \begin{table}[H]
    \centering
    \scalebox{0.8}{
    \begin{tabular}{|c|c|c|c|}
    \hline
    $C$ & $D$ & $D^\perp$ & $(C \star D)^\perp$ \\
    \hline \rowcolor{Gray}
    $[256,1]$ & $[256,37]$ & $[256,219,8]$ & $[256,219]$ \\  \hline
    $\mathbf{[256,1]}$ & $\mathbf{[256,30]}$ & $\mathbf{[256,228,8]}$ & $\mathbf{[256,228]}$ \\
    \hline
    \end{tabular}}
    \caption{First row: RM-based construction; second row: subfield subcode of $J$-affine variety code.}
    \label{table:comparisonS(j-affine)andRM1}
    \end{table}

     The PIR rate of the scheme based on the first row is $\frac{219}{256}$, while  the rate of the one based on the codes in the second row  is $\frac{228}{256}$. This  shows an improvement of the new construction over the one based on Reed--Muller construction.

    \item For $r = 8$ and $q = 2$, (i.e. $n_J = 512$),  the involved codes have the following parameters.

    \begin{table}[H]
    \centering
    \scalebox{0.8}{
    \begin{tabular}{|c|c|c|c|}
    \hline
    $C$ & $D$ & $D^\perp$ & $(C \star D^\perp$ \\
    \hline \rowcolor{Gray}
    $[512,1]$ & $[512,46]$ & $[512,466,8]$ & $[512,466]$ \\  \hline  
    $\mathbf{[512,1]}$ & $\mathbf{[512,34]}$ & $\mathbf{[512,478,8]}$ & $\mathbf{[512,478]}$ \\
    \hline
    \end{tabular}}
    \caption{First row: RM-based construction; second row: subfield subcode of $J$-affine variety code.}
    \label{table:comparisonS(j-affine)andRM2}
    \end{table}

   The PIR rate of the scheme based on the first row is $\frac{466}{512}$; whereas the rate of the one based on the codes in the second row rises to  $\frac{478}{512}$, again outperforming the construction based on  Reed Muller codes.

\end{itemize}
\end{example}

\subsection{Comparison with Berman Codes}  
In this final section, we compare the construction based on subfield subcodes of $J$-affine variety codes with a PIR scheme derived from Berman codes~\cite{BermanCodes}, which are defined over the binary field and are transitive. Berman codes were first introduced in \cite{berman1}, and they strictly include Reed-Muller codes as a subfamily. Reed–Muller codes can be recursively constructed using the $(u \mid u+v)$ construction \cite{Huffman_Pless_2003}, and, in an analogous manner, Berman codes can also be defined recursively through a construction similar to the $(u \mid u+v)$ method \cite{berman2,berman3}.

We consider the same setting as in Example~\ref{exp:binary.n49}, with the following parameters:
\[
q = 2, \quad m = 2, \quad q^3 = 8, \quad n_J = 49.
\]
Let \( D \) be the code defined by
\[
\Delta_D = \{(0,0), (1,0), (2,0), (0,1), (0,2)\} \cup \{(4,0), (0,4)\},
\]
so that \( d(D^\perp) \geq 4 \), ensuring a privacy level of \( t = 3 \).

Now consider the defining set \( \Delta_C = \{(0,0)\} \). In this case, the dimension of $(C \star D)^{\perp}$  is \( 42 \). This construction results in a higher PIR rate compared to the scheme based on Berman codes~\cite{BermanCodes}, in which the storage code \( C = \mathrm{DB}_7(0,2) \) has parameters \([49, 1]_2\), and the retrieval code \( D = \mathrm{DB}_7(1,2) \) has parameters \([49, 13]_2\). Both schemes share the same storage rate \( R_s = 1/49 \) and privacy level \( t = 3 \). 

Specifically, our scheme achieves a PIR rate of \( 42/49 \), while the Berman code-based scheme attains a PIR rate of \( 36/49 \). In the table below, the bolded rows highlight the parameters obtained via our construction, whereas the shaded row corresponds to the scheme based on the duals of Berman codes. For the same storage rate and privacy level, our construction offers a superior PIR rate.

\begin{table}[h]\begin{center}\begin{tabular}{|c|c|c|c|c|c|c|c|}\hline$ C $  & $D$ & $D^{\bot}$ & $C\star D$ & ${(C\star D)}^\bot$ &$R_{S}$ & Privacy & $R_{PIR}$  \\\hline$\mathbf{[49,1]_2}$  & $\mathbf{[49,7,21]_2}$&$\mathbf{[49,42,4]_2}$ & $\mathbf{[49,7]_2}$ & $\mathbf{[49,42]_2}$ &$\mathbf{1/49}$ &$\mathbf{3}$ & $\mathbf{42/49}$ \\ \hline
$\mathbf{[49,1]_2}$  & $\mathbf{[49,10,20]_2}$&$\mathbf{[49,39,4]_2}$ & $\mathbf{[49,10]_2}$ & $\mathbf{[49,39]_2}$ & $\mathbf{1/49}$& $\mathbf{3}$ & $\mathbf{39/49}$ \\ \hline  \rowcolor{Gray}
$[49,1]_2$  & $[49,13,16]_2$&$[49,36,4]_2$ & $[49,13]_2$ & $[49,36]_2$ &$1/49$ & 3 & $36/49$ \\ \hline
\hline\end{tabular}
\end{center}
\caption{ Comparison of Berman codes-based scheme (shaded rows) with $J$-affine variety codes-based scheme (boldface rows).}\label{table:berman}
\end{table}

\begin{table}[h]\begin{center}\scalebox{0.5}{\resizebox{\columnwidth}{!}{\begin{tabular}{l|l}\hline$ \Delta_C $  &  $\Delta_D$    \\ \hline$   \{(0,0)\}$ & $\{(0,0),(1,0),(2,0),(0,1),(0,2),(4,0),(0,4)\} $   \\$   $ &  $\{(0,0),(1,0),(2,0),(0,1),(0,2),(4,0),(0,4),(1,1),(2,2),(4,4)\} $  \\\end{tabular}}}\end{center}\caption{Cyclotomic cosets used for   codes in Table \ref{table:berman}. }\label{table:VI}\end{table}

\begin{remark}
    In general, when selecting the storage code $C=\mathrm{DB}_n(r_C,m)$ and the retrieval code $D=\mathrm{DB}_n(r_D,m)$ as given in \cite[Table 1]{BermanCodes}, and setting $ m = 2 $  with $(r_C, r_D) = (0,1) $, the resulting scheme has $ t = 3 $, a storage code rate of $ R_{s} = \frac{1}{n^2} $, and a retrieval rate of $ \frac{(n-1)^2}{n^2} $. Specifically, the parameters of the involved codes are as follows: 
\[
C: [n^2, 1, 49], \quad D: [n^2, 13, 7], \quad D^{\perp}: [n^2, 36, 4]. 
\]

For comparison reasons, when choosing $n = q^s - 1 = 2^s - 1 $, to achieve the same privacy level and storage rate while obtaining a better retrieval rate, our scheme must satisfy the conditions 
\[
\dim((C\star D)^{\perp}) = \dim(D^{\perp}) > (n-1)^2 \]
\[n^2-\dim(D^{\perp}) < n^2-(n-1)^2=2^{s+1}+3 \]
\[\dim(D) < 2^{s+1}-3.
\]
Since we fix $t=3$, we know that $\Delta_{D^{\prime}}
= \{(i,j) | (i+1)(j+1)<4 \}$, so $\#\Delta_{D^{\prime}}=5$. Thus, if we have  $$\#\left(\cup_{P\in\Delta_{D^{\prime}}}I_P\right) < 2^{s+1}-8, $$ i.e., $\#\Delta_D < 2^{s+1}-3 $, then $\dim((C\star D)^{\perp}) > (n-1)^2$, which gives better retrieval rate than the scheme described in \cite{BermanCodes}. 
\end{remark}

\section*{Acknowledgements}
The authors thank R. San-Jos\'e for helpful comments on this paper. They also thank the reviewers for their careful reading and insightful feedback, which have significantly improved the manuscript.

\ifCLASSOPTIONcaptionsoff
  \newpage
\fi

\bibliographystyle{plain}

\bibliography{schur.bib}

\end{document}